\pdfoutput=1
\documentclass[acmsmall]{acmart}
\AtBeginDocument{%
  \providecommand\BibTeX{{%
    \normalfont B\kern-0.5em{\scshape i\kern-0.25em b}\kern-0.8em\TeX}}}

\usepackage{sidecap}
\usepackage{lineno,hyperref}
\usepackage[cache=false,draft=true]{minted}
\usepackage{xcolor}
\usepackage{placeins}
\usepackage{array}
\usepackage{float}
\usepackage{multirow}
\usepackage{arydshln}
\usepackage{enumitem}
\usepackage{amsmath}

\usepackage{amssymb}
\usepackage{pifont}
\usepackage{listings}
\usepackage{graphicx}
\usepackage{sidecap}
\usepackage[caption=false]{subfig}
\newcommand{\cmark}{\ding{51}}%
\newcommand{\xmark}{\ding{55}}%
\definecolor{ORN2}{rgb}{0.54,0.2,0.14}
\newcommand{\textpck}[1]{\texttt{\textcolor{ORN2}{ #1}}}
\setlist[enumerate]{nosep}

\def\frameworktool{\texttt{DiVA}}
\def\expandname{{\bf Di}ffusion \textbf{V}isualization and \textbf{A}nalysis}  
\newcommand{\framework}{\frameworktool}
\newcommand{\frameworkexpand}{\expandname}
\modulolinenumbers[5]



\begin{document}
\title{DiVA: A Scalable, Interactive and Customizable Visual Analytics Platform for Information Diffusion on Large Networks}
\author{Dhruv Sahnan}
\affiliation{%
  \institution{IIIT Delhi}
  \country{India}}
\email{dhruv18230@iiitd.ac.in}
\author{Vasu Goel}
\affiliation{%
  \institution{IIIT Delhi}
  \country{India}}
\email{vasu18322@iiitd.ac.in}
\author{Sarah Masud}
\affiliation{%
  \institution{IIIT Delhi}
  \country{India}}
\email{sarahm@iiitd.ac.in}
\author{Chhavi Jain}
\affiliation{%
  \institution{IIIT Delhi}
  \country{India}}
\email{chhavi19117@iiitd.ac.in}
\author{Vikram Goyal}
\affiliation{%
  \institution{IIIT Delhi}
  \country{India}}
\email{vikram@iiitd.ac.in }
\author{Tanmoy Chakraborty}
\affiliation{%
  \institution{IIIT Delhi}
  \country{India}}
\email{tanmoy@iiitd.ac.in}

\begin{abstract}
With an increasing outreach of digital platforms in our lives, researchers have taken a keen interest in studying different facets of social interactions. Analyzing the spread of information ({\em \textit{aka}} diffusion) has brought forth multiple research areas such as modelling user engagement, determining emerging topics, forecasting the virality of online posts and predicting information cascades. Despite such ever-increasing interest, there remains a vacuum among easy-to-use interfaces for large-scale visualization of diffusion models. In this paper, we introduce \textbf{\framework} -- \frameworkexpand, a tool that provides a scalable web interface and extendable APIs to analyze various diffusion trends on networks. \framework\ uniquely offers support for simultaneous comparison of two competing diffusion models and even the comparison with the ground-truth results, which help develop a coherent understanding of real-world scenarios. Along with performing an exhaustive feature comparison and system evaluation of \framework\ against publicly-available web interfaces for information diffusion, we conducted a user study to understand the strengths and limitations of \framework. We noticed that evaluators had a seamless user experience, especially when analyzing diffusion on {\em large} networks.
\end{abstract}

\maketitle
\section{Introduction}
\label{sec:intro}
\setlist[itemize]{noitemsep, topsep=0pt}
Information diffusion is one of the core areas in the study of social networks, both online and offline. A diffusion starts with the sender/seed node spreading the information to its neighbours, who further spread it to their neighbours, and so on. Thus, a diffusion model achieves its task by identifying a path or a tree (branches representing the transmission), capturing the evolution of information over time. On a larger scale, since communities are interlinked, this information can also jump from one community to another. While in the real world, epidemic models are being successfully applied to the study of dynamics of disease outbreaks \cite{10.2307/2529951}, only recently have they been utilized to study the spread of information on the digital platforms \cite{7029011}.
 
Diffusion dynamics is used to understand the flow of information, diseases, content \cite{ContentSecurity}, identifying behaviour patterns of influential users \cite{InfluenceUsersSurvey}, etc. Despite a growing body of work that focuses on mathematical modelling of information diffusion, little work has been attempted to accommodate visual analysis of diffusion dynamics \cite{socialwave, opflow}. Recently, there has been tremendous development in analyzing epidemics, and their spread due to the COVID-19 pandemic \cite{corona-tracker,10.1145/3430984.3430989, PanMTIYJ20, YangXPJ20}. \textit{However, due to their niche application and platform dependency, the tools mentioned above cannot be easily exported to work for different social platforms and settings.} In addition, while all major technology companies have access to an extensive network of users and resources, the visualization supporting these networks is usually in-house\footnote{\url{https://www.ivysys.com/social-network-analysis}}, and close-sourced\footnote{\url{https://cambridge-intelligence.com/keylines/}}. Moreover, \textit{there is a limited number of publicly-available interfaces to simulate and visualize information diffusion on large networks.}

To overcome the shortcomings mentioned earlier, in this paper, we introduce \textbf{\framework} (\frameworkexpand)  --- an open-source, customized web interface for the study of information diffusion. The proposed tool allows users to study complex social networks visually in the form of a clustered node-link structure. It makes it easier to study numerous diffusion patterns (spread/adoption) of posts, news articles, blogs, music, social trends, epidemic outbreaks, political campaigns and public opinion. \framework\ presents the diffusion analysis in three different forms --- (a) statistics, (b) plots, and (c) an interactive diffusion visualization over the network. While the first two are macroscopic analysis, the last feature allows for the microscopic study of information diffusion at a community (or even node) level and at specific timestamps. Note that the focus of this paper is to provide a salable and easy-to-use system for studying the existing diffusion models rather than proposing another information diffusion algorithm. \emph{The most unique and promising feature of \framework\ is its ability to visually and numerically compare the simulations of two diffusion algorithms at the same time. Hence by extension, it supports simultaneous comparison of the results of a diffusion algorithm against the real-world (a.k.a, ground-truth) diffusion patterns.} This allows the users to analyze the changes in diffusion trends when models or parameters are modified and can help practitioners determine the best course of action under varying settings. Furthermore, keeping the needs of fellow researchers in mind, \emph{\framework\ allows the user to simulate their custom algorithms on the platform and analyze how they fare in a real-world network as well.} Finally, to provide the holistic experience of working with social networks, \framework\ also supports the traditional statistical network measures such as PageRank, node clustering, degree distribution, etc.

We provide a detailed comparison of the features of \framework\ with other available applications and point out how \framework\ fares much better than any of these tools. Although there are a few tools and packages available to visualize diffusion, there are none with a comparable set of features and performance exhibited by \framework.

\textbf{Contributions.} In short, our major contributions are as follows:
\begin{enumerate}
    \item \framework\ is the first visualization platform that allows users to visually compare two diffusion models or one diffusion model with the ground-truth diffusion pattern.
    \item \framework\ provides a highly customized environment to the users --- it allows them to upload custom diffusion algorithms and networks for analysis.
    \item \framework\ provides several metrics to explain the diffusion pattern of a model with downloadable reports.
    \item Unlike existing tools/platforms, \framework\ is highly scalable. We perform both qualitative and quantitative evaluations to test the same.
\end{enumerate}


\textbf{A demo video of the tool as well as codebase of \framework\ is publically available at  \href{https://github.com/LCS2-IIITD/DiVA}{\underline{\textcolor{blue}{link}}}.}

\section{Background and Related Work}
\label{sec:related_work}
In the past decade, social media platforms have been on exponential rise. Data availability on platforms like Twitter, Facebook, and Reddit makes it feasible to understand online user behaviour. While there are many studies and tools for visual analysis of user behaviour, little work has been done for visualizing information diffusion, simulated or otherwise. To the best of our knowledge, \framework\ is the first tool that provides a highly scalable and customizable platform for visualizing information diffusion models. 

\textit{\textbf{Analyzing User Interaction.}}  Some existing tools aid in the analysis of social interaction amongst users. This information can be beneficial in tasks like community detection, social recommendations, influential user detection, etc. These tools generally combine computational aspects like graph theory, network analysis, demographic analysis \cite{10.1145/3272973.3273001}, ethnography, and epidemiology to provide users with important semantic and cluster information. Tools such as Vizster \cite{vizster}, iOLAP \cite{iolap}, and NodeXL \cite{nodexl} use edge grouping algorithms to create clusters of nodes. Some algorithms such as hierarchical edge bundles \cite{hierarchical,hierarichalshareflow}, geometric based edge clustering \cite{geometry}, and motif clustering \cite{motif}, use network properties. Incorporating semantic information in clustering algorithms such as semantic substrates \cite{semanticsubs}, top-N node filtering \cite{nodefilter}, and probabilistic topic modeling \cite{probmodeling} helps analyze social interaction better. 

\textit{\textbf{Analyzing Social Media Content.}} Many tools add to these features by providing content analysis. Tools like FireAnt \cite{fireant}, Visual-VM \cite{visualvm}, SAVIZ \cite{saviz}, EpiGrass \cite{epigrass} and Google+ Ripples \cite{ripples} allow a content-based analysis of the network. Data like geo-tags, though sparse, are useful in analyzing mobility patterns. Geographical data is also helpful in analyzing user stance, behavior, and knowledge across the world. Other tools like FireAnt \cite{fireant}, SocialFlow \cite{socialflow}, HistoryFlow \cite{historyflow} and Whisper \cite{whisper} also provide topic analysis. Extracting the posts' terms, hashtags, and topics helps with opinion, product, and marketing analysis. This is also useful to study information diffusion in blogs and entertainment sites \cite{blogvisual2020,musicanalysis,videodiffusion}. Chen et al. \cite{dmap} introduced an extensive user-central visualization and analysis tool for social media data through graphs.
On the one hand, some browser extensions \cite{BRENDA, pop, TweetAbuse} are made available to run on top of social media platforms, providing platform-specific analysis of users, content flow, and even content moderation \cite{10.1145/3272973.3274056}. Meanwhile, some web interfaces like Reddit-Network-Vis\footnote{\url{http://whichlight.github.io/reddit-network-vis/}} that focuses on visualizing the top comments of a nested Reddit post, or Topic-Flow \cite{Malik2013TopicFlowVT}, and TweetViz \cite{tweetviz} which focus on visualizing the temporal evolution of a topic on Twitter, etc. have also been proposed. Changing the platform or the content format will lead to a significant change in these tools' capabilities. We, thus, need a tool that can help us visualize information spread irrespective of the underlying social media platforms.

\textit{\textbf{Analyzing Information Diffusion.}} While many studies visualize social media data, very few tools support information diffusion algorithms. NDlib \cite{ndlib} is a Python package for analyzing information diffusion algorithms. It also contains a basic module that visualizes and compares multiple diffusion algorithms on a randomly generated graph. EpiModel \cite{epimodel} is an R package that provides epidemic diffusion models. Built on top of NDlib and EpiModel are the \texttt{NDlib-Viz} and \texttt{Epinet} respectively. While both these visualization modules provide algorithm simulation, \texttt{Epinet} provides a more robust and customized tool than \texttt{NDlib-Viz} which has a bare minimum visualization interface. \framework\ provides a much more vast array of features and customization than these tools. RECON\footnote{RECON Projects \href{https://www.repidemicsconsortium.org/projects/}{https://www.repidemicsconsortium.org/projects/}} provides a collection of R packages for visualization and analysis of outbreak data. In Section \ref{sec:eval}, we compare \framework\ with these tools in more detail.

\begin{table}[!t]
\centering
  \caption{Some existing tools/libraries for visualizing large networks.}
  \label{tab:RWgraph}
  \scalebox{0.95}{
   \begin{tabular}{p{0.165\linewidth}|p{0.13\linewidth}|p{0.64\linewidth}}
    \hline
    \textbf{Name} & \textbf{Type} & \textbf{Features}\\\hline
    \href{tableau.com}{\underline{\textcolor{blue}{Tableau}}}  & Tool & Business analytics tool; supports input from excel and database systems; includes a large collection of visualizations and dynamic geographical data analysis.\\
    NodeXL \cite{nodexl} & Tool & Complex network visualization and analysis tool; supports input through excel files. \\
    Gephi \cite{gephi} & Tool & Open source network visualization and analysis software; supports multiple input formats; extensive support for visual customisation. \\
    NetworkX \cite{networkx} & Python Library & Library to perform extensive statistical analysis on complex networks with basic visualizations. \\
    \href{https://socnetv.org}{\underline{\textcolor{blue}{SocNetV}}} & Tool & Open source social network visualization and analysis tool. \\
    \href{https://ggplot2.tidyverse.org}{\underline{\textcolor{blue}{gplot2}}} & R Package & Statistical and graph visualization; easier to create and customise multi-layer graphics.\\
    \href{https://d3js.org}{\underline{\textcolor{blue}{D3.js}}} & JavaScript Library & Open source library highly used for data driven work; used extensively for data visualizations; primarily uses SVG and HTML Canvas; has limited support for WebGL in network visualization. \\
    Cytoscape \cite{Otasek2019} & Tool & Open source complex network visualization library; provides options for multiple layouts and customizable  appearance.  \\
    \href{https://js.cytoscape.org/}{\underline{\textcolor{blue}{Cytoscape.js}}} & JavaScript Library & .js library for the tool Cytoscape; supports dynamic and interactive networks and can be integrated with other apps for network analysis and development. \\
    \href{https://www.fusioncharts.com}{\underline{\textcolor{blue}{FusionCharts.js}}} & Javascript Library & Basic graphs as well as network visualizations; also has maps of world and countries; extensive charts library; it is a paid tool. \\ 
    \href{https://pyviz.org}{\underline{\textcolor{blue}{PyViz}}} & Python Library & Open source collection of packages for complex network visualization; supports data manipulation and exporting for easier integration with common data science libraries.\\
    DMap+ \cite{10.1145/3183347} & Tool & Visulise the user and event centric information spread on Weibo.\\
    E-Map \cite{8585638} & Tool & Utilizes map-like visualizations to analyse and study the patterns in social media event data.\\
    McVA \cite{CHEN201819} & Tool & A visualization tool to study multiple hierarchical dataset of pesticide usage across different geographies.\\
    Aureole \cite{10.1007/s12650-017-0467-x} & Tool & Another domain-specific visualization tool to study hierarchical dataset of cellular networks.\\
  \hline
\end{tabular}}
\end{table}

\textit{\textbf{Visualizing Large Networks.}} Analysing large social networks with millions of nodes requires agile visualization tools. Early works used C/C++ libraries to create fast visualizations \cite{pajek,askgraphview,sonia}. With the development in computational hardware and other languages, many libraries are available for fast, clutter-free, and easy network visualization.
While DyNetVis \cite{10.1145/3019612.3019686} is one of the open-source, free, and actively developed projects in the area of dynamic and large-scale visualization, it is a stand-alone tool, unlike the web-based options this study explores. Even though DyNetVis provides an excellent set of tools for understanding the temporally evolving networks, it has limited support for diffusion visualization algorithms. Table \ref{tab:RWgraph} lists some of the popular complex network visualization tools/libraries. Some of them facilitate network analysis but hardly support diffusion analysis. Some works \cite{largevizgraph2d,largegraph} tried to visualize high dimensional data by reducing it to 2D similar to t-SNE \cite{tsne}. Nevertheless, they create static graphs, which do not help analyze diffusion. Some of the hierarchical visualizations can also be useful for large-scale networks, where the connections can be unrolled at varying levels. A combination of techniques like Louvain \cite{Blondel_2008}, or PansyTree \cite{9086199} can be helpful here.

\textit{\textbf{Visualizing Information Diffusion.}} Techniques like flow maps, heat maps, word clouds, and projection matrices have been used to study social networks to help in performing a comparative analysis (both pair-wise and aggregated) of the information. Chen et al. \cite{10.1145/3183347} proposed a method that takes force-directed atlas and community-level information as input and studied the \{ego, event\}-centering visualization on top of the incoming Weibos. On Similar lines, E-Map \cite{8585638} was introduced as a tool to heuristically build a map-like visualization of the various topical keywords under discussion on social media. It uses keyword affinity and resharing behavior to construct a spatial-temporal map of the topics. While both the tools are adequate for low-scale unique nodes to analyze the information crisply, no comments were made on the scalability and performance of the systems. Moreover, the systems are not open-sourced. However, the diffusion analytic provided by DMap+ and E-Map can be used as an extension layer for \framework, since the former requires the input to be a diffusion cascade. In another study, an interactive tool \cite{7192688} was developed to explain the movement of people using geotags and hashtags information coupled with the time of posting the social media content. Such interactive visualizations first require the scraping and storing of social media content before being analyzed. We do not provide a data scraping mechanism. However, once preprocessed in the form of a network, our tool can be utilized to run simulations. One can also study the ground-truth user-movement patterns with cities as nodes and diffusion algorithms defining users' movements among the city nodes. Being content-agnostic, \framework\ can be adapted to study diffusion patterns of various use cases.

For diffusion visualization, a dynamic and customized visualization library is required. In a web browser environment, we find that using WebGL\footnote{WebGL API: \url{https://developer.mozilla.org/en-US/docs/Web/API/WebGL_API}} backed libraries renders the network faster and results in smoother animations than other libraries.
Meanwhile, libraries like NepidemiX\footnote{NepidemiX: \url{http://nepidemix.irmacs.sfu.ca/}} (outdated now) and Outbreak2\footnote{Outbreak2: \url{http://www.repidemicsconsortium.org/outbreaker2/}} (a part of RECON projects) only provide statistical analysis like the growth charts, unlike \framework\ which provides an animation-like visualization of the diffusion process. Epicontacts (RECON package) supports large graph visualization (not diffusion). However, overall, the projects are very distributed, do not provide an integrated analysis tool, and lack many features introduced in \framework. Other tools like Epigrass \cite{epigrass}, and GLEAMviz \cite{van2011gleamviz} focus on epidemic and outbreak simulations. Epigrams support custom simulations but fall short in the visualization aspect. GLEAMviz is an advanced tool that supports multiple simulation algorithms and geographical visualization. However, it cannot be employed for social networks. 

\textit{\textbf{Comparative Visualization.}} Diffusion cascades can be represented in a tree-like manner, owing to which techniques that study the formation and analysis of hierarchical data can also be extended to study information diffusion. Tree-visualization for single and multi-attribute structures has been an area of active research \cite{doi:10.1057/ivs.2009.29}. Employing the ArcTrees \cite{ArcTrees} or the BarCodeTree \cite{8845772} mechanism could be one way of visualising hierarchical information. Even though these techniques claim to be salable, it is not easy to convert an interconnected network of entities into disjoint cascades and flat hierarchies as a single node can get influenced by multiple neighbors. Meanwhile, tools like McVA \cite{CHEN201819} and Aureole \cite{10.1007/s12650-017-0467-x} are built with niche applications in mind. The former supports the visualization of multiple hierarchical datasets for studying pesticide usage. The latter is a tool for cellular networks represented on hierarchical maps. These tools have a specific data prepossessing pipeline that can be difficult to extend to different social media formats. Similar to our proposed tool, the comparative visualization tools are interactive and allow users to perform a range of filtering and exploration. As most real-world data is sparse, incomplete, and intractable, having interactive visual systems helps translate the user's knowledge (especially domain expertise) to navigate and enrich the data. This is specially useful in understanding route trajectories \cite{7850970}, and characterising DNA sequences \cite{10.1093/nar/gkz404, Shah2005}. 

\section{Motivation and Target Audience} Our primary motivation for developing this tool is a lack of an easy-to-use, scalable and interactive interface for diffusion visualization. The results of diffusion algorithms are hard to explain, given the complex interactions underpinning them. Visualizing these results on the network at different timestamps and comparing them against the aggregated metrics can help better understand how the information is gradually spreading. With \framework, practitioners can effectively communicate their findings to a broader range of audiences. Combining different forms of information and presenting them in different visual formats enables the user to cherry-pick the level of granularity and reporting desired by them. For example, analyzing the raw results of a diffusion model is not easy for a non-expert user. Meanwhile, manually visualizing the complex sequence/traversal of information is untraceable even for expert users. 

As a customized and open-source tool, our primary target audience is the research community that can quickly check existing or custom diffusion methods' viability or perform the ground-truth analysis. This analysis can range from studying computer viruses spread over a network to real viruses over a population. Practitioners within the research community and stakeholders like law enforcement and news media can use \framework\ to study the dissipation of hate speech or fake news and test control strategies for the same. The dual diffusion visualization acts as a visual source for A/B testing of two similar (but differently tuned) or contrasting hypotheses/policies to compare their impact. It can help policymakers decide on the best way forward for applications from vaccination schema to content marketing. Additionally, researchers working on mathematical modeling of newer diffusion systems can dry run the new model's robustness against baselines even on large networks. While for the initial user studies of \framework, we started with a pool of participants that more closely resembled researchers. We will be surveying other stakeholders for future iterations and feature integration. 

\subsection{User Requirement Survey}
In order to better understand the limitations of the current system and address these commonly-occurring pain points in our proposed system, we conducted an anonymous user survey without revealing any information about our tool. A group of $25$ participants ($19$ males and $6$ females) consisting of researchers, engineers, and epidemiologists, were recruited. The participants had varying levels of expertise in network analysis and diffusion. There were $10$ experts,  $9$ for moderate expertise, and $6$ novice participants. Even though all $25$  participants understood network analysis, $18$ of them had actively worked on network diffusion. The complete set of survey questions is provided in Appendix \ref{app:req_survey}. Based on the survey of existing tools used for analyzing the spread of information, two graph visualization tools, Cytoscape \cite{Otasek2019} and Gephi \cite{gephi}, came up as the most frequently used tools. Apart from these, some expert users also mentioned about \texttt{Epinet} and \texttt{NDlib-Viz} \cite{ndlib}. One user also enlisted QGIS\footnote{\url{https://qgis.org/en/site/}} as a tool for overlaying geospatial data on top of diffusion visualizations. Even though both Cytoscape and Gephi (JVM-based desktop tools) are frequently used for network analysis, the users complained about their bulky setup and lack of support for large networks.
Interestingly, $16$ users opted for a web-based application when asked about the platform-application setup preference. Even though there are famous desktop tools, users prefer a web-based application with less time to set up. Consequently, we focus on web-based interfaces used for diffusion visualizations for our comparative study and performance evaluation. For our use case these interfaces are \texttt{Epinet} and \texttt{NDlib-Viz}.

For the survey question \textit{"Can you list out some limitations that you may find while using some of the said tools?"}, the top pain points of the users came up to be:
\begin{itemize}
    \item (C1): Lack of support for large networks.
    \item (C2): Lack of support for different graph-input formats.
    \item (C3): Resource and memory-intensive tools are hard to set up.
    \item (C4): Lack of scriptability and customizability, and less interactive UI.
\end{itemize}
Subsequently, when asked \textit{"According to your use cases, can you list three most important features that you feel must be present in any network diffusion visualization tool that you may use?"}, the top most desired features included:
\begin{itemize}
\item (F1): Support for large networks.
\item (F2): Easy to use, customized visualization of diffusion.
\item (F3): Saptio-temporal analysis of the information flow.
\item (F4): Availability of key network and diffusion statistics at a glance.
\item (F5): Ability to save and load checkpoints.
\end{itemize}
Combining the information gained from the survey, we were able to see how well our proposed system tackled some of the pain points listed by the users. We were also able to map our feature sets with the top feature requests of the users. 
\begin{itemize}
    \item (S1): As evident from the system architecture (Section \ref{sec:arch}) and performance evaluation (Section \ref{sec:perf_eval}), we build a robust system to support large scale networks. 
    \item (S2): A two-pronged human evaluation of \framework\ (Section \ref{sec:user_eval}) points out its ease of use and interactiveness. We especially incorporated tooltips and default parameter values.
    \item (S3): Enlisted in Section \ref{sec:customization} are various customization options supported by \framework\ -- a variety of input graph formats, support to test user-defined algorithms, and input seed nodes. Apart from this interface customization is also supported in terms of updating node/edge colors. Once a simulation has been run, the node color of different classes of a diffusion model can also be updated.
    \item (S4): To bridge the gap between the visualization of information flow and its corresponding statistical information, we provide a mechanism to view the animated diffusion and glance at the fine-grained and network-level statistics. Additionally, the plots for the rate of information spread and the results of the simulations are made available for download and reuse.   
\end{itemize}
The user requirement survey not only helped us better understand the feature set of\framework, but also pointed out some interesting features for future work (enlisted in Section \ref{sec:conclusion}).

\section{Major Challenges}
We curate these challenges based on the literature study on large-scale diffusion visualization and initial iterations of building our system. We also highlight how these challenges shaped our design decisions. While the major of challenges boil down to engineering and design decisions, there are a few research-oriented challenges as well.
\subsection{Engineering Challenges}
\begin{itemize}
\item \textit{Closed-sourced desktop tools}: Network visualization tools are often closed-source and licensed as desktop applications. Testing such tools for performance is difficult (and reverse engineering can even be illegal). Additionally, desktop-based tools often face issues with installation due to incompatibility among libraries and versions on different operating systems. Talking specifically about Gephi \cite{gephi}, one needs to increase the default JVM memory allocated to the application to work with large graphs. Additionally, on Linux, one may have to uninstall and reinstall a specific JDK version for OpenGL\footnote{\url{https://www.opengl.org/}} to work seamlessly for large networks. The alternative is a web-based app, where HTML can seamlessly embed WebGL\footnote{\url{https://www.khronos.org/webgl/}}, and we can take advantage of today's browser to support visualizations effectively. On the other hand, OpenGL has far more capabilities than WebGL. Consequently, after reviewing the existing desktop tools and visualization libraries, we opted for a web-first approach to avoid dependency hassles and cross-platform compatibility. Based on the organization's use case, one can locally or publicly host \framework. Meanwhile, secured session management is handled via Google Authentication.
\item \textit{Inactive open-source tools}: The other major issue is lack of maintenance of open-source tools and libraries like LargeViz \cite{largeviz} and \texttt{NDlib-Viz}. LargeViz was built to support visualizations for large graphs but is no longer maintained. The last commit on LargeViz happened in 2016\footnote{\url{https://github.com/lferry007/LargeVis/commits/master}}. A similar fate has befallen \texttt{NDlib-Viz} -- a visualization web tool designed to run and analyse diffusion models supported by its parent library \texttt{NDlib} \cite{ndlib}. While \texttt{NDlib} is still actively maintained, the visualization library to attract non-experts and non-programming users fell short of its purpose by not providing active documentation support. The last commit on \texttt{NDlib} was in 2018\footnote{\url{https://github.com/rinziv/NDLib_viz/commits/master}}. While we included the \texttt{NDlib-Viz} in our comparative study, we found that the front-end could not support a random graph of more than $10k$ nodes. As discussed in detail in Section \ref{sec:perf_eval}, we see the choice of embedding framework impacted \texttt{NDlib-Viz}'s ability to load and visualize large graphs. We had to run a few iterations to reach the combination of WebGL and three.js that proved best for our use case (details of the system architecture are provided in Section \ref{sec:arch}).
\item \textit{Handling large graphs}: In a user survey on challenges in graph processing \cite{10.1145/3186728.3164139}, scalability and visualization were reported as top challenges when dealing with large graphs. Loading large graphs is resource-intensive; we look at $O(n^2)$ relationships to map in the worst-case scenario, where $n$ is the number of nodes. It also causes an issue with the readability of large graphs, as quickly adjusting layouts is not possible. Various layout mechanisms have been proposed and examined \cite{Kwon2018} to increase the readability of large graphs. In the case of \framework, since we aim to look at the spread of information at both macro and microscopic levels, dimensionality reduction-based techniques were not considered. We initially started with layouts that supported node features. However, it quickly became apparent that visualizing node features on the main canvas was not feasible for large graphs. We shifted this functionality under the node viewing component, where the information of the currently clicked/selected node can be viewed, and this information dynamically changes upon clicking another node. More than the node feature, the underlying graph structure does the heavy lifting of determining how the information will flow with the network. Thus, by abstracting node-level information, we could map the network with edge lists which are faster to process. We use force-directed based layout optimization. D3.js has out-of-box support for this layout. By default, it helps preserve the community structure and ensure that adjacent nodes are as close as possible. Lastly, we realized that we could decrease the reloading time by having the layout coordinates precomputed for the next session by allowing users to save the loaded graph (details of the system architecture are provided in Section \ref{sec:arch}).
\item \textit{Level of modularity}: R Epidemics Consortium (RECON)\footnote{\url{https://www.repidemicsconsortium.org/projects/}} provides a suite of packages covering various aspects of analyzing and visualizing epidemic models. However, to use different functionalities, different libraries need to be imported. Instead of opting for such a breakdown of our application, we decided to keep various aspects intact to provide both networks and diffusion level analytics and visualization under one hood. Once running, the users are not required to set up or integrate any additional resources and stay within the browser for a consistent user experience. It should be noted that the code structure is modular, with API endpoints defined for different execution units.
\item \textit{UI Design decisions}: It took a few iterations to determine the placement of various components within \framework's UI. While we did not have the resources to run exhaustive UI testing to determine the best placement of the components, the features were clubbed based on the functionality they provided. The default panels were set based on the core usage of simulating a diffusion model and visualizing its output. Additionally, instead of keeping the interface as a single-page web application (like \texttt{NDlib-Viz}), we provide separate browser tabs for the single and dual visualization modes. When users click on dual visualization, they are taken to a new browser tab. It frees users to run multiple simulations on the same graph separately and download their results. In hindsight, our two-step user study (Section \ref{sec:user_eval}) highlights that the user finds the design framework intuitive and easy to follow.
\end{itemize}

\subsection{Research Challenges}
\begin{itemize}
    \item \textit{Test multiple hypotheses}: Research work often requires testing of multiple hypotheses to determine the best course of action. Researchers and practitioners often tweak different system parameters and want to visually and numerically capture how the tweaking impacts the flow of information within the network. This is the major motivation behind developing the dual visualization mode. While \texttt{NDlib-Viz} does support running multiple models in the same session,  users can view the network and plot information for only one model at a time.
    \item \textit{Experiments with the newer version of a diffusion model}: While a portion of research work involves running a comparison against existing baselines/standard diffusion models, researchers also want to test how their proposed diffusion algorithms perform. In order to facilitate the same, the interface needs to be fluid and support scripting (Python in our case). Apart from researchers, practitioners and policymakers also require running comparisons against ground-truth information, and being able to upload that onto the system is what \framework's custom scripting supports.
    \item \textit{Reproducible and extendable results}: A significant focus of research and policy decisions is being able to go back and retrieve the results of a hypothesis. Additionally, using the obtained results, the users should be able to extend upon it (send to a secondary dashboard, automate archiving of reports, etc.). In order to encourage reproducibility and extensibility, we support:
    \begin{itemize}
        \item Downloading the network's statistical analysis in a .diva format. This is particularly useful when the users are running experiments on random graphs. Additional reproducibility for random graphs is adjusted via the seed parameter that we allow the user to set while generating a random graph in \framework.
        \item The diffusion iteration results returned by the NDlib backend are also available for downloading. Using multiple such results for different experiments, users can perform a granular analysis of where and how different models converge and diverge.
        \item Diffusion plots (for both visualization modes) can be saved as pdfs.
    \end{itemize}
\end{itemize}

\section{Interface Overview}
\label{sec:divis}
\framework\ offers two main diffusion visualization modes --- {\bf Primary visualization mode}, and {\bf Dual visualization mode}. Additionally, the system offers a wide array of features to analyze the network provided by the user. This section explores the visualization modes and their utilities. An end-to-end workflow of the interface is enlisted in Appendix \ref{app:end2end_flow}.

\begin{figure}[!ht]
    \includegraphics[width=\textwidth]{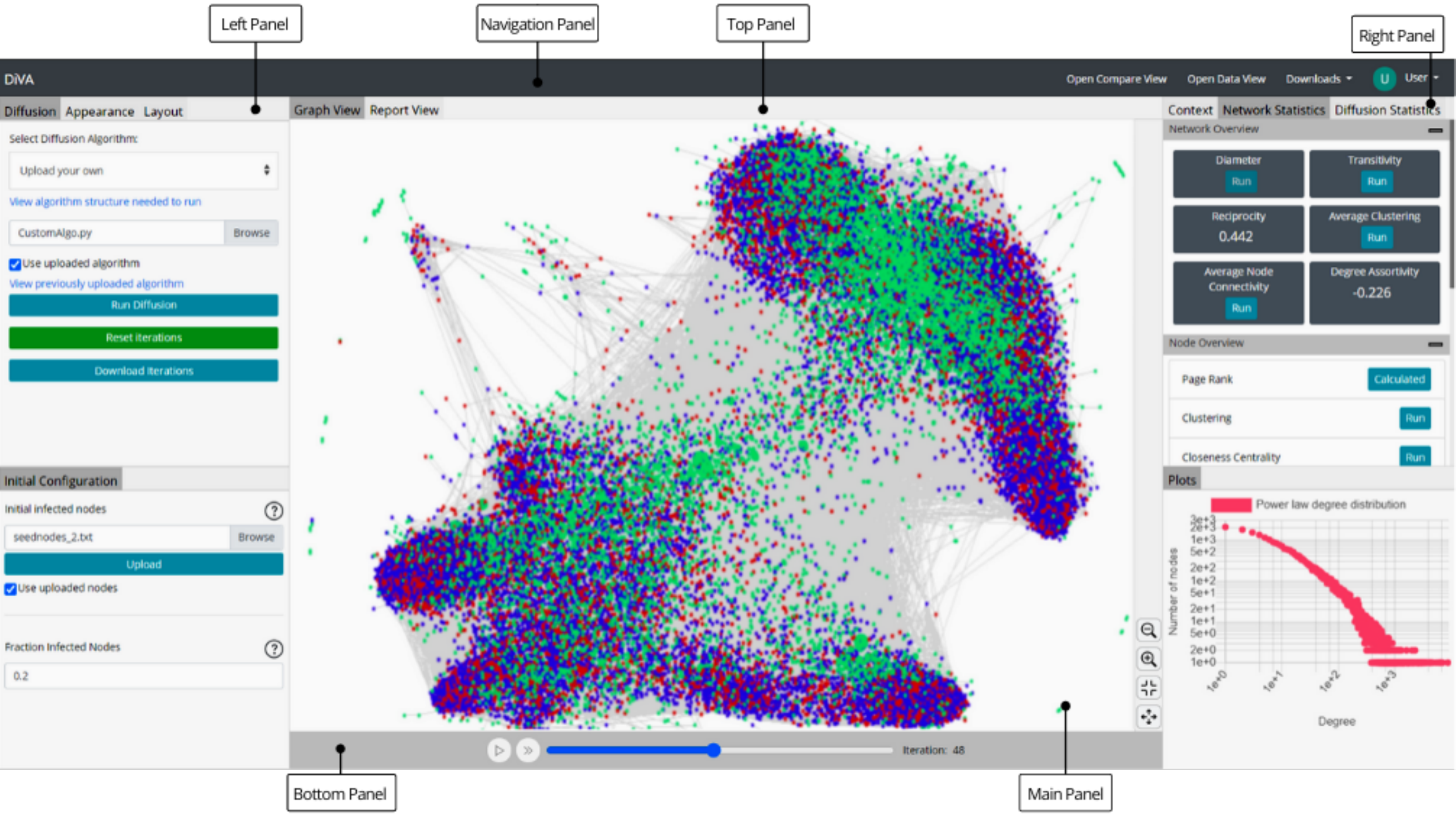}
    \caption{Overview of \framework\ for the Primary visualization mode at iteration $t=48$. The user uploaded a network consisting of $25K$ nodes and $700K$ edges. The user also uploaded a custom algorithm (\texttt{CustomAlgo.py}) and provided a list of initially infected nodes (\texttt{seednodes\_2.txt}). The layouts of the six panels visible for this mode are separately marked. The Main Panel loads the central canvas containing the network. The panels on both sides of the main canvas contain the tools necessary to perform the diffusion analysis. The panels on the top and bottom act as the navigation bar and media control panel, respectively.}
    \label{fig:diva_single_viz}
\end{figure}  

\subsection{\textbf{Primary Visualization Mode}}
\label{sec:primary_viz_mode}
\begin{figure*}[!h]
    \includegraphics[width=0.97\columnwidth,keepaspectratio]{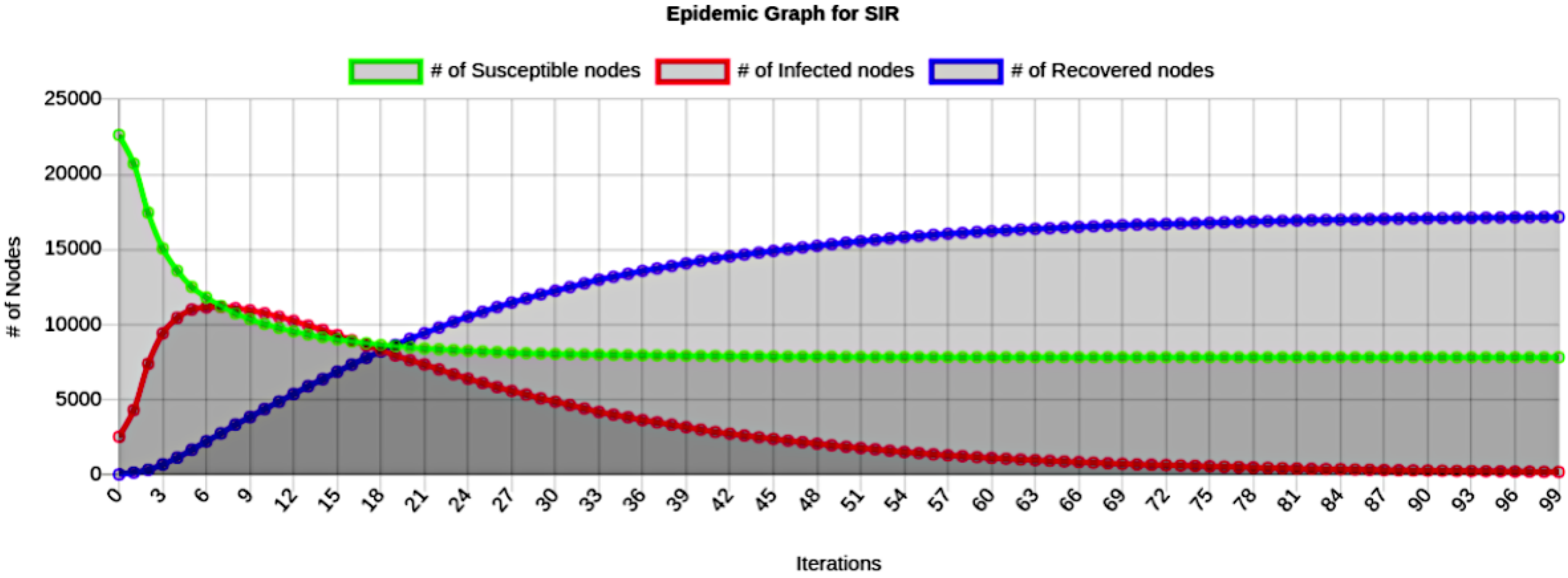}
    \caption{Example of \texttt{Report View} for SIR model simulation in the Primary visualization mode.}
    \label{fig:single_report_view}
\end{figure*}

By default, this mode is loaded by the system, as illustrated in Figure \ref{fig:diva_single_viz}. At the core of \framework\ lies its ability to visualize large-scale networks on a web-based interface. Once a user sets the initial network, \framework\ displays it in the main panel where a node's color is graded according to its degree. The darker the node's color, the higher its degree of centrality. After submitting a diffusion query, the user can graphically view the spread of the diffusion model. Once the graph view is triggered, it dynamically updates the color of a node as per the node status at the given iteration. An intentional delay is added to help the user grasp the node colorization change between two iterations. The interactive interface allows the user to check out the diffusion snapshot (visually and numerically) at a particular timestamp. As depicted in Figure \ref{fig:single_report_view}, a diffusion trend of the SIR diffusion model displays the count per class among the three classes (infected, susceptible, and removed) per iteration.
\subsubsection{\textbf{Utility of Primary Visualization Mode.}} The platforms that support network-level statistics rarely support diffusion analysis and vice-versa. \framework, on the other hand, offers a unique set-up bringing the two together. It is only natural that while analyzing the diffusion patterns of a network, the user may also be interested in running network-level statistics. For example, a user interested in simulating the information spread starting at the most influential nodes can get the Page Rank information of the nodes from the \texttt{Data View} and provide a custom seed list accordingly.  The main aim of dynamically visualizing the diffusion model is to help the users better understand the depth and breadth of the spread of the contagion entity. It additionally allows the user to check out the diffusion snapshot at a particular timestamp and even pause it. As the slider is moved backward/forward in timestamps, the graph appearance and diffusion statistics are updated accordingly. Apart from the visualization, the \texttt{Diffusion Statistics Tab} (\ref{fig:right_panel}.C) lists down the node count per class per iteration, updating as the iterations proceed. The users can either download the results of diffusion iterations as a raw JSON file or save the PDFS of diffusion trends from the \texttt{Report View Tab}.

\subsection{\textbf{Dual visualization Mode}}
\label{subsec:dual_diff}
\begin{figure*}[!ht]
    \includegraphics[width=\columnwidth,keepaspectratio]{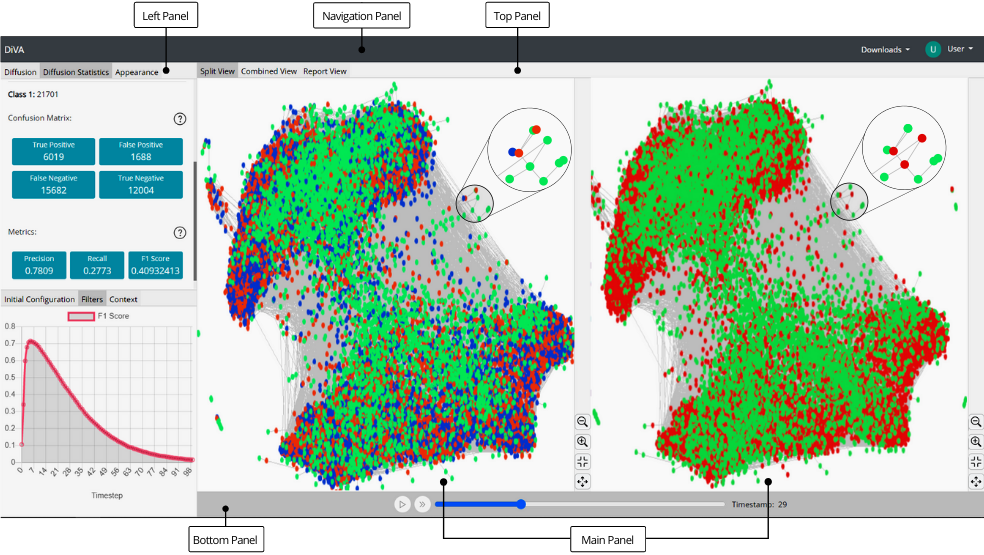}
    \caption{Overview of \framework\ for the dual diffusion visualization mode at iteration $t=29$. The user uploaded a network of $25K$ nodes and $700K$ edges. The layouts of the five panels visible for a dual diffusion visualization are additionally marked. SIR and SIS models are selected for comparison. Diffusion results for SIR and SIS appear on the left and right splits of the main panel, respectively. Green, red and blue being the susceptible, infected and recovered nodes respectively. One can observe even the minor differences in the diffusion algorithms. We have highlighted one such example by showing a zoomed up version of a cluster of nodes, in which different nodes are infected for both the SIR and SIS algorithms.}
    \label{fig:diva_compare}
\end{figure*}

\begin{figure}[!t]
    \centering
    \includegraphics[scale=0.43,keepaspectratio]{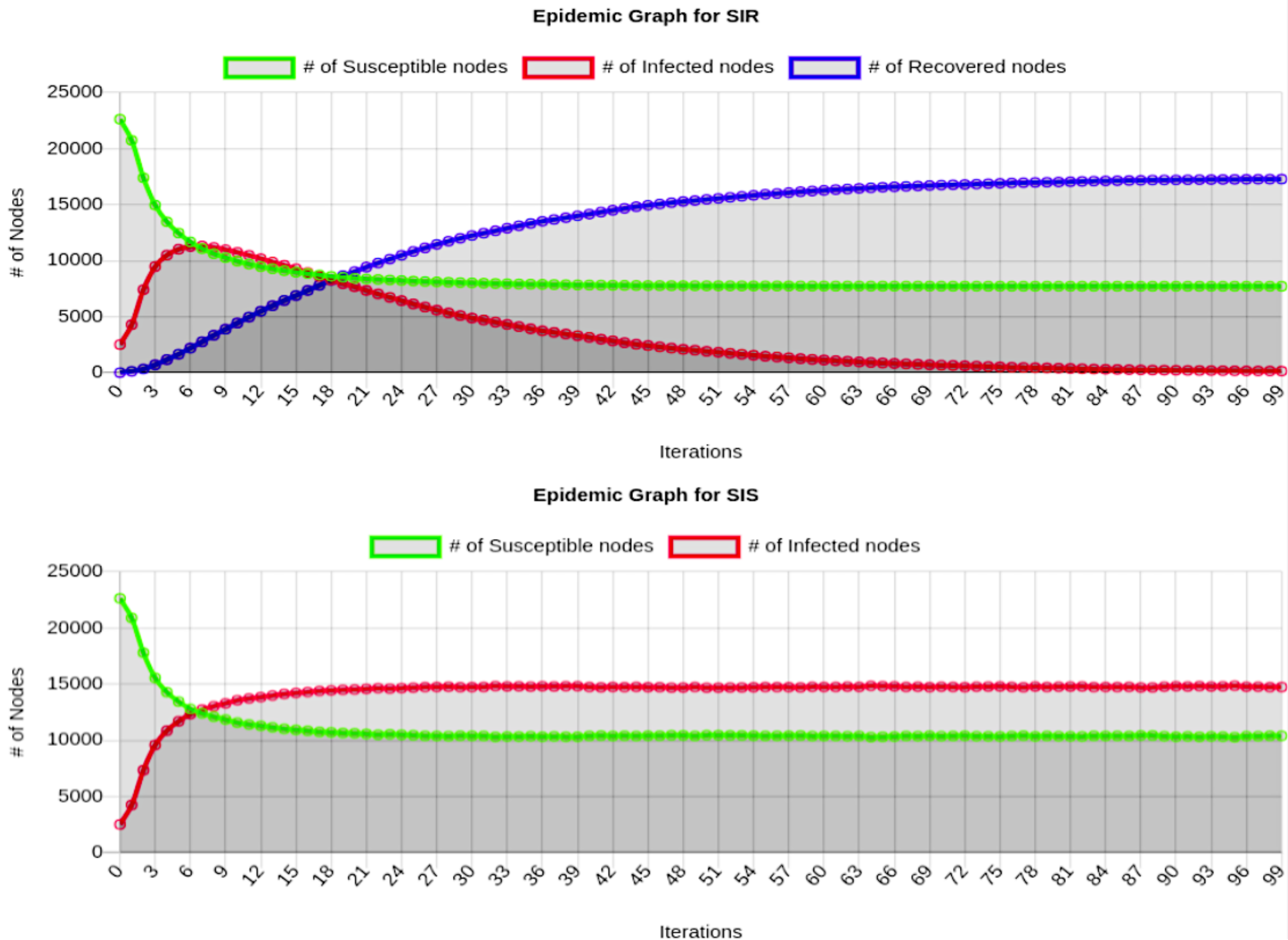}
    \caption{Example of \texttt{Report View} for SIR and SIS model simulations in the Dual visualization mode.}
    \label{fig:dual_report_view}
\end{figure}

In \framework, we introduce an advanced diffusion visualization view, wherein a user can compare two different diffusion algorithms in a split view on the \textbf{Main Panel} which is illustrated in Figure \ref{fig:diva_compare}. 
As stated before, \framework\ aims to understand the breadth and depth of a diffusion process visually. It becomes even more necessary in the dual diffusion mode as more than one moving component is involved. The same number of iterations and initially infected node sets are used for both models. It enables a one-to-one mapping of the results per iteration can be obtained. Users can visualize the diffusion in one of the formats:
    \begin{itemize}
        \item \texttt{Split View}: As the name suggests, here, the Main Panel is split into two, with the same graph appearing in the two splits. The two splits update the node color of the respective algorithms, with the second diffusion algorithm treated as a ground truth. \emph{The main aim here is to view how a change in the model or its parameters causes a change in the rate of spread.} 
        \item \texttt{Single View:} Here, the results of both the simulations are visualized on a single graph. Each model updates the nodes as per the corresponding status for each iteration. If two models operate on the same node, then the second model's status (treated as ground-truth by default) prevail. \emph{The main aim of the single view mode is to provide a visual understanding of the convergence and divergence of the results obtained from the two models.} 
    \end{itemize}
The user can switch between the two modes at any point in time, as both the views share the common bottom panel, i.e., the same iteration-number mapping. The \texttt{Report View} now contains three set of plots: (a) Diffusion trends one per algorithm- Figure \ref{fig:dual_report_view}. (b) A plot of  F1-score per iteration. (c) A plot of commonly infected nodes per iteration. If a node's status is set as infected by both the models for a given iteration, it is counted towards the commonly infected nodes for that iteration.
Figure \ref{fig:diva_compare} shows a snapshot of dual diffusion comparison in action for two different diffusion models in \texttt{Split View} format. The left is the SIR (susceptible-infected-recovered) model, and the right is the SIS (susceptible-infected-susceptible) model. Imagine these simulations to be a study of chickenpox vs. common cold. In the case of the former, a person usually remains immune after recovery (SIR), whereas in the case of the latter, one can always catch a cold (SIS). Naturally, one would expect that a system infused with a recovery setup will eventually have lesser infected nodes than a system with no recovery. Upon zooming in on the top-right corner of the network in Figure \ref{fig:diva_compare}, we can observe this phenomenon locally. With green, red and blue being the susceptible, infected, and recovered nodes, we observe in the zoomed-in section 2 infected nodes in the SIR model compared to 3 in the SIS model. Consequently, by zooming into various regions within a large network, the users can navigate the spread of contagion at a more granular level for different timestamps. Meanwhile, on the \texttt{Left Panel} the aggregated class-wise statistics per timestamp are available.

\subsubsection{\textbf{Utility of Dual Visualization Mode.}} It supports three different comparison set ups:
\begin{itemize}
    \item \textbf{Intra-model Comparison}: The user may choose to compare the results with different hyper-parameters of the same algorithm. An example is showcased in case study. (Section \ref{sec:case_study})
    \item \textbf{Inter-model Comparison}: This allows the user to compare two different algorithms. Users may also use this function to evaluate a custom algorithm against a baseline model of their choice. For example, when comparing the impact of the spread of malware in a router network, the analysts can compare the spread with or without blocker nodes.
    \item \textbf{Ground-truth Comparison}: \framework\ also provides an interesting functionality of simply visualizing the diffusion using the ground-truth results. For example, if a policymakers want to compare the impact of their proposed model against a currently implemented policy, they can utilize the current ground results to compare with the proposed simulated results.
\end{itemize} 

\section{Interface Customization}
\label{sec:customization}
\begin{figure}[H]
    \centering
    \subfloat[]{\includegraphics[width=0.3\textwidth]{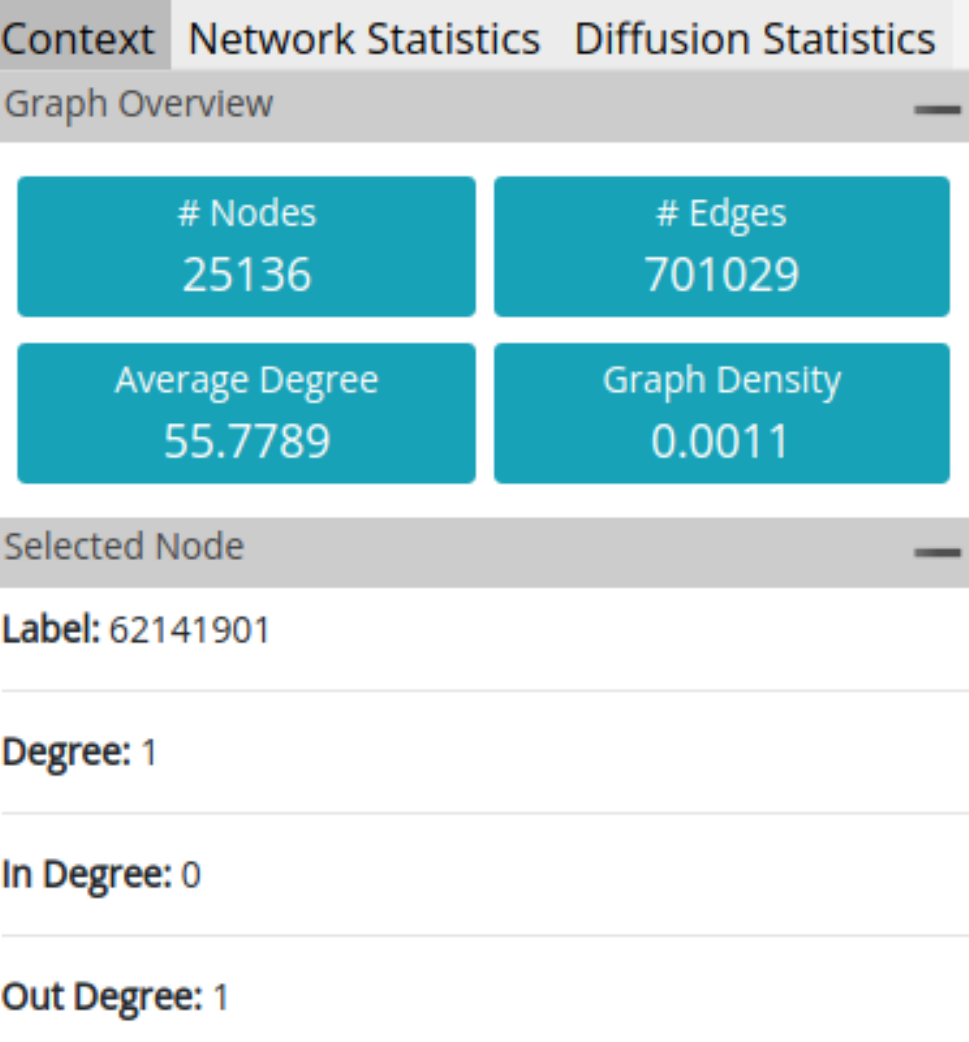}}\hspace{5mm} 
    \subfloat[]{\includegraphics[width=0.3\textwidth]{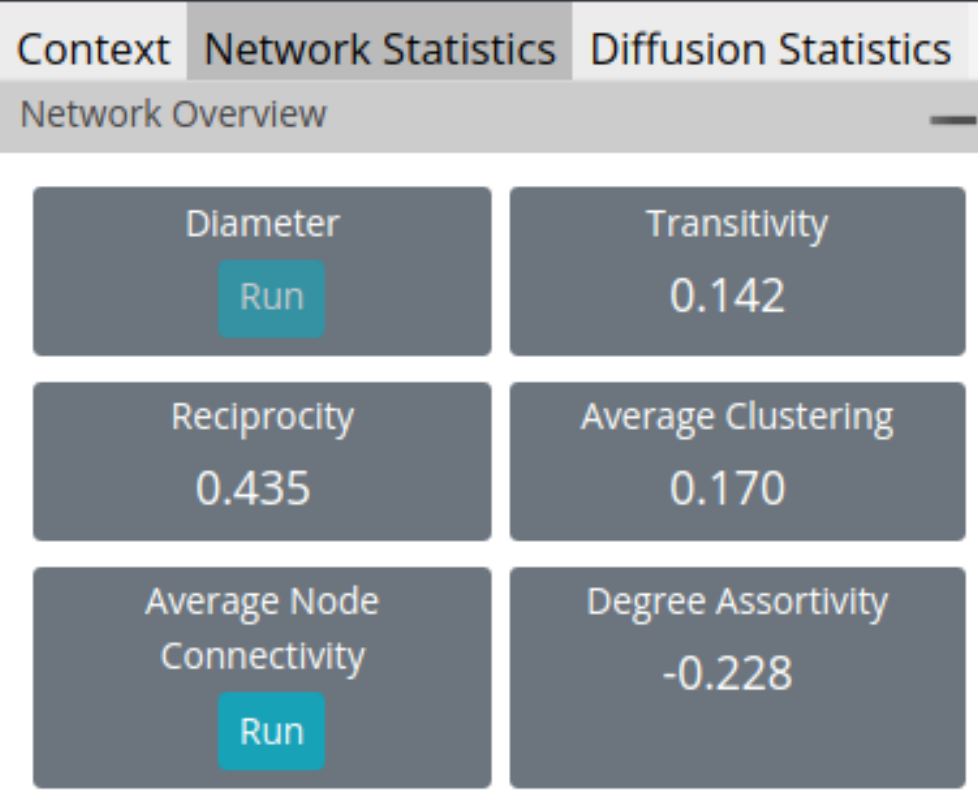}}\hspace{5mm} 
    \subfloat[]{\includegraphics[width=0.3\textwidth]{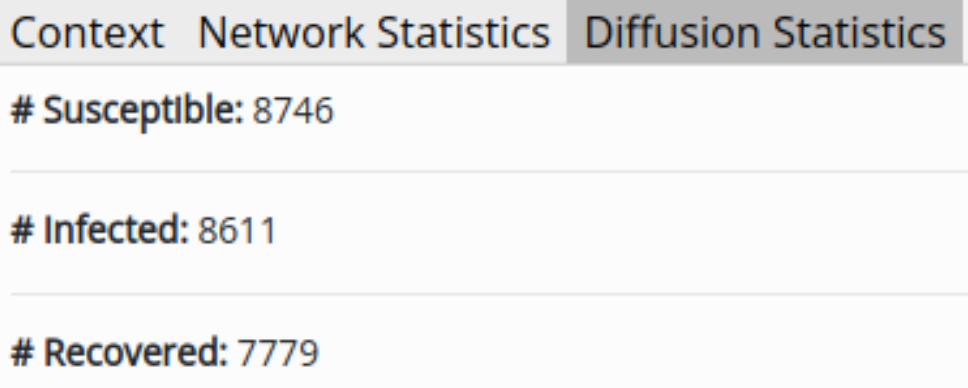}}
    \caption{Various components of the Right Panel: (a) \texttt{Context Tab}: Provides basic network information. (b) \texttt{Network Statistics Tab}: Showing the network level statistics. (c) \texttt{Diffusion Statistics Tab}: Showing the diffusion statistics}
    \label{fig:right_panel}
\end{figure}

\begin{figure}[H]
    \centering
    \subfloat[]{\includegraphics[width=0.3\textwidth]{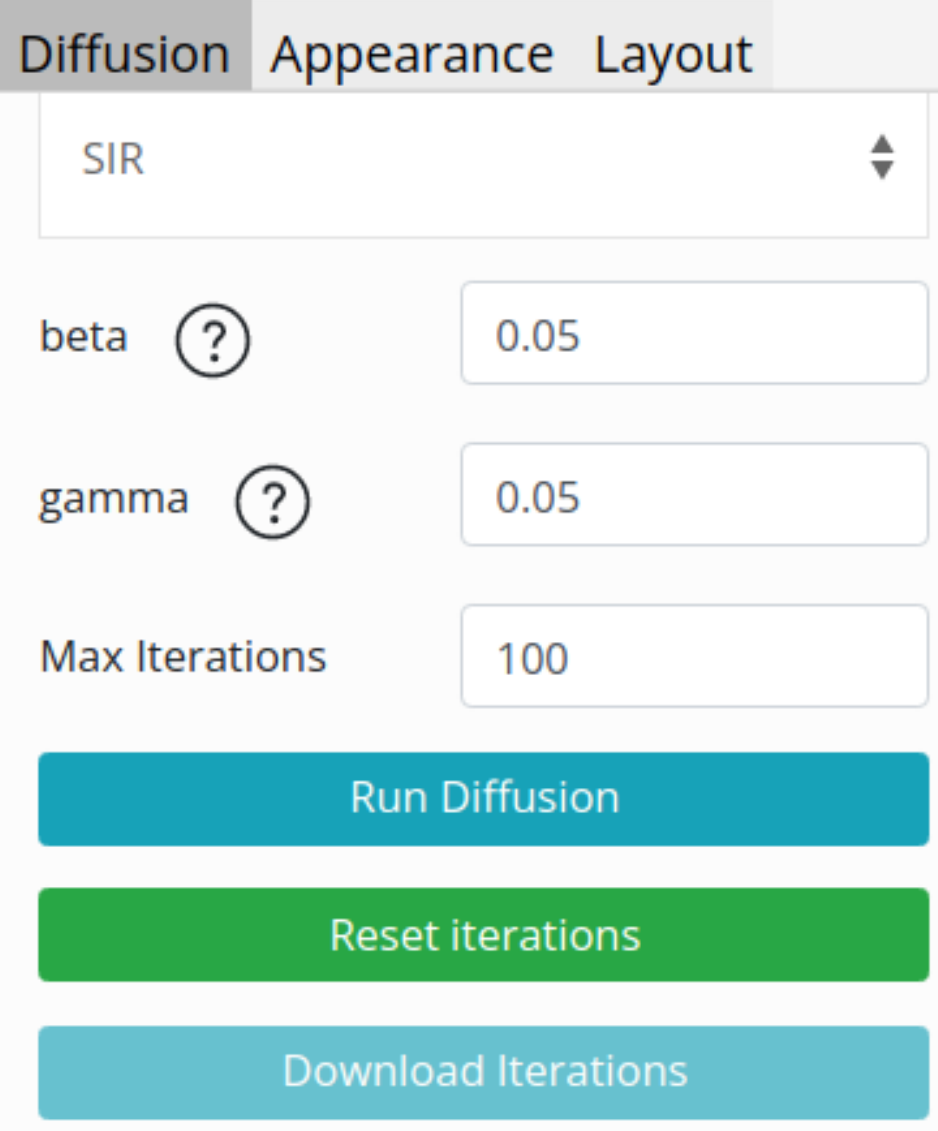}}\hspace{5mm} 
    \subfloat[]{\includegraphics[width=0.3\textwidth]{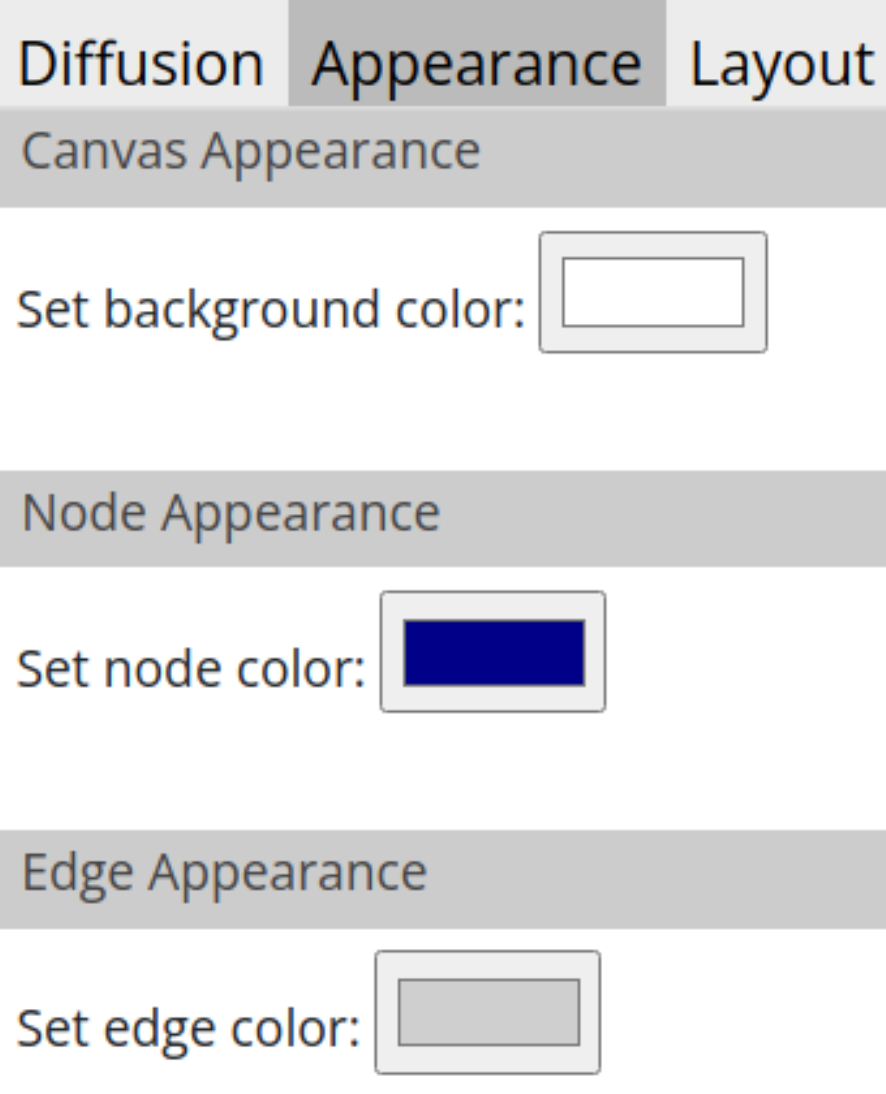}}\hspace{5mm} 
    \subfloat[]{\includegraphics[width=0.3\textwidth]{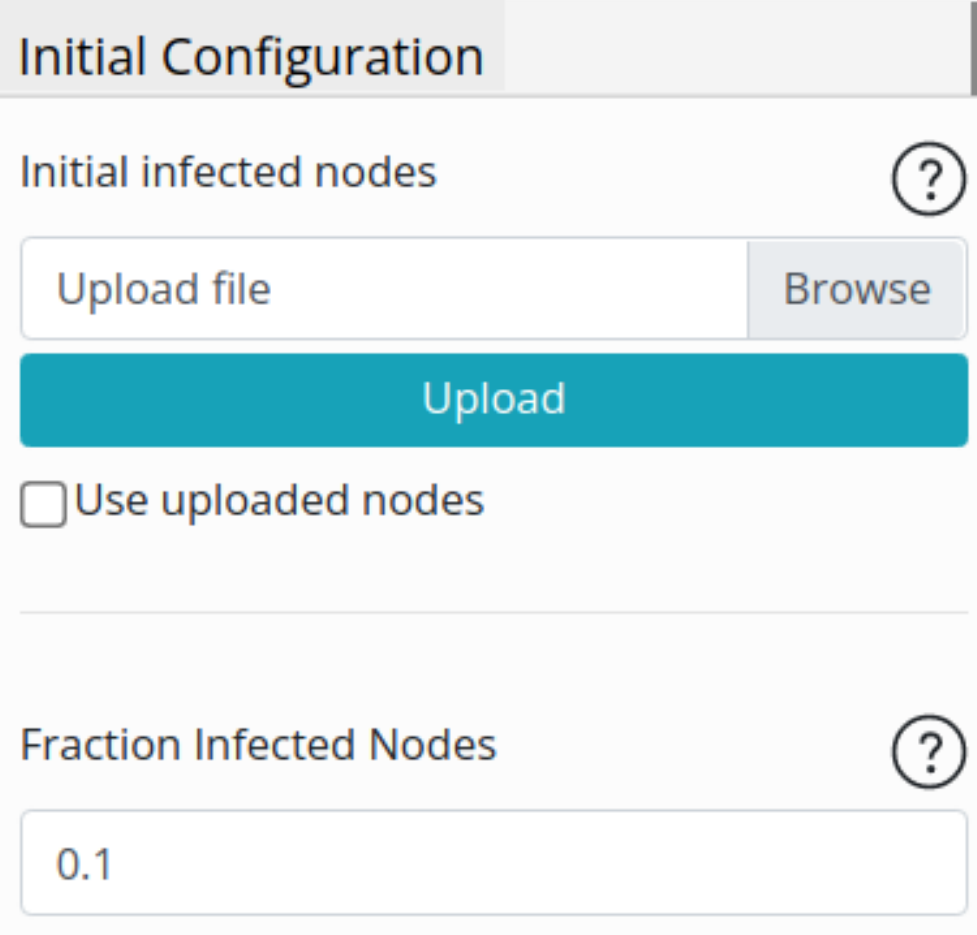}}
    \caption{Various components of the Left Panel: (a) \texttt{Diffusion Tab}: Selecting the SIR algorithm along with its parameters. (b) \texttt{Appearance and Layout Tab}: The default color combinations for the network. (c) \texttt{Initial Configuration Tab}: Setting $0.1\%$ of nodes as initially infected.}
    \label{fig:left_panel}
\end{figure}

Here, we highlight the customization offered by \framework:
\begin{itemize}
\item {\em Initial network}: Apart from the ability to create a random Endros-Renyi network to work with, \framework\ allows users to upload a custom network in one of the networkx supported formats (Appendix \ref{app:input_graph}).
\item {\em Diffusion algorithm}: Besides the wide range of epidemic algorithms supported \framework, we let the users upload their diffusion algorithms to test. Support of custom diffusion algorithm makes \framework\ accessible for various real-world use-cases. The format of custom algorithm is available in Appendix \ref{app:Custom_algo_code}. Additionally, the full list of epidemic models supported by \framework\ are enlisted along with their parameters in Appendix \ref{app:EP}.
\item {\em Diffusion parameters}: Once users select the diffusion algorithms, they can specify the values for the various diffusion parameters (else, default is used). Additionally, we provide the functionality for the users to upload a custom set of infected seed nodes (Figure \ref{fig:left_panel}.a and \ref{fig:left_panel}.c).
\item {\em Appearance}: Users can customize the color of the nodes and edges of the uploaded network.Additionally, once a diffusion simulation is run, the color that various classes of the diffusion model can take can also be updated (Figure \ref{fig:left_panel}.b). While the users cannot change the underlying network layout, they can rotate and zoom in/out. 
\item {\em Network statistics}: Users can run any available network statistics at a node/graph level (Figure \ref{fig:right_panel}.b), these results are also available under the \texttt{Data View Tab} (Appendix \ref{app:data_view}).
\end{itemize}

\section{System Design}
\label{sec:arch}
\begin{figure*}[!t]
    \includegraphics[width=\textwidth]{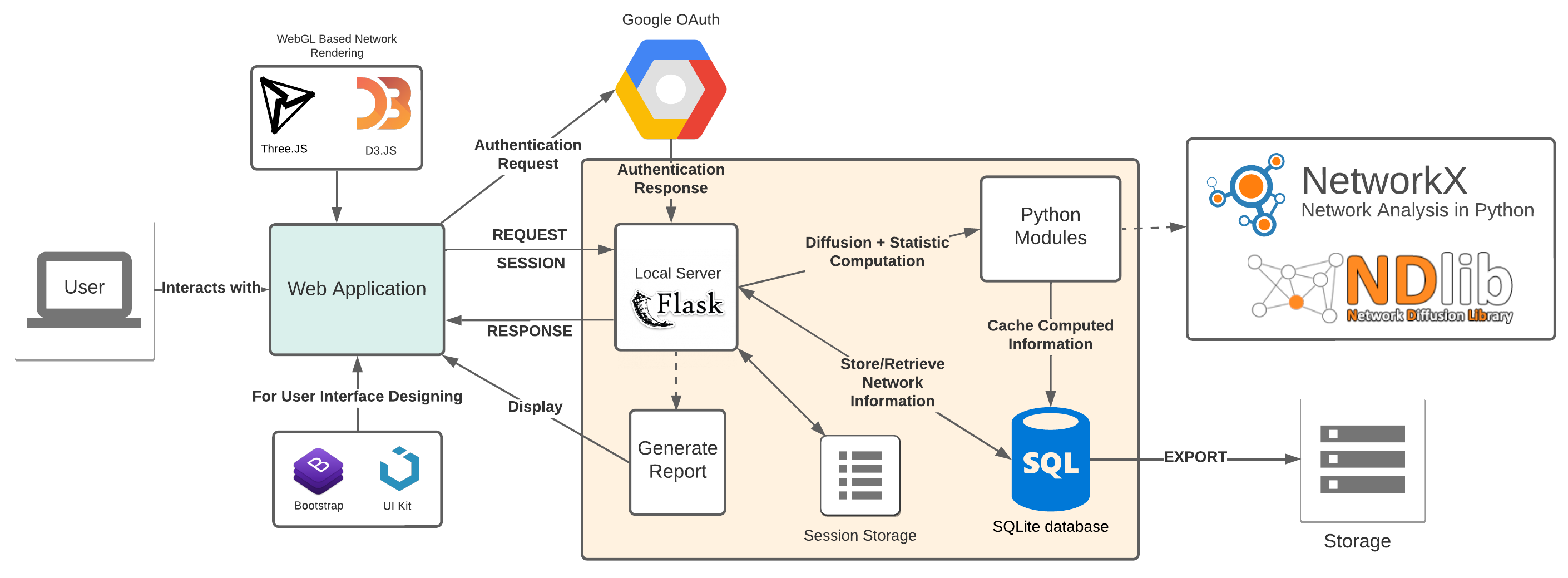}
    \caption{Architectural overview of \framework\ showcasing the interaction among various technical components.}
    \label{fig:diva_archt}
\end{figure*}

This section provides a breakdown of various technologies used to support different components of \ the framework. 
Being a web application, \framework\ is implemented using front-end technologies, namely JavaScript and HTML5 and back-end technologies including Flask\footnote{Flask: \url{https://flask.palletsprojects.com/}} (a microservice framework written in Python) and SQLite\footnote{SQLite: \url{https://sqlite.org/}} (a SQL database) for storage. 
The architectural overview of the web interface that supports \framework\ is provided in Figure \ref{fig:diva_archt}. We use Google’s authentication system and file system-based session management, which let multiple users work on \framework\ in parallel when used as a deployed application. As \framework\ is meant to be an online deployed tool, secure authentication and session management are necessary for the system.

\subsection{User Interface}
\label{sec:design_visual}
For the visualizations, we rely on the advancement of modern browsers and their increasing computational power to provide a cross-platform tool that is not limited by the user's operating system. 
The structure and layout of the tool are primarily managed with our custom CSS and JavaScript code. For the basic user-interface elements, we rely on UIkit 3\footnote{UIkit 3: \url{https://getuikit.com/}} and Bootstrap 4\footnote{Bootstrap 4: \url{https://getbootstrap.com/}}. 
The front-end and back-end communicate using AJAX and REST APIs.
\subsection{Visualizing Large Networks}
\label{sec:visualize_network}
Due to large file sizes, visualizing and analyzing large networks can tax the processor and memory. Creating a dynamic network visualization requires more processing and memory than static visualizations. The storage, file sharing, and analytical processes must be done carefully to avoid bottlenecks and blocked processes.

{\textbf{Back-end.}}
Within the Flask server, we use NetworkX for (a) modeling the network into an optimized data structure, (b) computing network statistics, and (c) making use of existing high-performance graph representation file formats. We chose NetworkX, given its ease of use and popularity. Although libraries such as graph-tool \cite{peixoto_graph-tool_2014} and igraph \cite{igraph} offer better performance in terms of computation, they require complicated setups. They do not offer equal support for different operating systems. 
While we plan to bring out integration for igraph for users looking to work with even more extensive networks; however, for networks with up to $100$k nodes, it has been observed that NetworkX is a viable solution.

We use a custom network representation file (a JSON-based edge list format) and stream it to the front-end asynchronously in chunks through a custom streaming API using JavaScript web workers. The size of the representation file can reach up to $46$ MBs for a network of $75$k nodes and $4.5$M edges, which, if transferred in one chunk, hogs the server's resources. Additionally, we provide a mechanism for the users to reuse the computed locations of nodes and edges by saving a custom \texttt{.diva} file, which stores a compressed version of all the computations run by the user. It allows the user to load a previously saved graph quicker, \emph{as the saved file lets us skip some initial steps that the server performs for setting up the network (effectively reduces the load time by a factor of five for larger graphs. Results reported in Table \ref{tab:diva_perf}).} The streaming mechanism helps reduce the server load, essentially enhancing the multi-user functionality of the system. 

{\textbf{Front-end}.}
The representation so obtained is fed into d3-force’s layout\footnote{d3-force-layout\url{https://github.com/d3/d3-force}} generation algorithm, which calculates node positions by modeling the network as a physical system of interacting particles, and simulates physical forces between the particles, resulting in a clustered node-link structure of the network. Being able to highlight different communities clearly, provides an easy way to study the diffusion process. Large-scale networks are loaded within a few minutes (see Table \ref{tab:diva_perf}) because of the method used to stream coordinates of nodes and edges from the server as computed by the force-directed algorithm. This system essentially results in a network topology that is quite similar to one that is generated by ForceAtlas2 \cite{forceatlas2}. Further, we use three.js\footnote{three.js:\url{ https://threejs.org/}} to plot these nodes and the corresponding edges onto a WebGL context, which is used for all network visualizations of \framework.  \emph{During the initial architectural analysis of existing web interfaces for diffusion, we found that systems using HTML5 Canvas or SVG contexts for network visualizations become unresponsive for larger graphs; case in point, the \texttt{NDlib-Viz}}. As depicted in Table \ref{tab:diva_perf}, we have been able to overcome this problem by employing a combination of WebGL and d3.js\footnote{ d3.js: \url{https://d3js.org/}}. 

\subsection{Network and Diffusion Analysis}
\label{sec:visualize_report}
\textbf{Back-end.}
The server also maintains a user’s network state and provides all the network and diffusion algorithms statistics. All network statistics are stored in the database and can be reused. The server calls are executed asynchronously, and the respective visual components are updated on the front-end with the response as generated by the server. For running diffusion algorithms, we use \texttt{NDlib} --- a network diffusion library. Additional, \framework\ supports custom algorithms (details provided in Appendix \ref{app:Custom_algo_code}). 
 
{\textbf{Front-end}.}
The additional visualizations of charts and plots for reporting statistics (as depicted in Figure \ref{fig:single_report_view}) are created using the JavaScript charting library, Chart.js\footnote{Chart.js: \url{https://www.chartjs.org/}}. The reports are generated entirely on the front-end and rely only on the server for pre-computed network statistics. Finally, we use DataTables\footnote{DataTables: \url{https://datatables.net/}} to list out the statistical information for all nodes in a tabular format.
    
\begin{table}[!t]
        \centering
            \caption{Feature comparison o three web-based diffusion visualization tools --- \texttt{Epinet}, \texttt{NDlib-Viz} and \framework.}
            \label{tab:feature_comp}
            \scalebox{0.75}{
            \begin{tabular}{p{0.5\linewidth}|p{0.1\linewidth}|p{0.11\linewidth}|p{0.08\linewidth}}
            \hline
            \textbf{Feature} & \textbf{\texttt{Epinet}} & \textbf{\texttt{NDlib-Viz}} & \textbf{\framework}\\
            \hline
            Web based application & \cmark & \cmark & \cmark \\
            \textbf{Extendable APIs} & \xmark & \cmark & \cmark \\
            Generate random network & \cmark & \cmark & \cmark \\
            \textbf{Upload custom network} & \xmark & \xmark & \cmark \\
            Network statistics (basic) & \cmark & \cmark & \cmark \\
            Network statistics (advanced) & \xmark & \xmark & \cmark \\
            Node information & \cmark & \xmark & \cmark \\
            Interactive graph display  & \xmark & \xmark & \cmark \\
            Basic graph controls (zooming/panning) & \xmark & \cmark & \cmark \\
            \textbf{Customize Graph Appearance} & \xmark & \xmark & \cmark \\
            \textbf{Custom diffusion model} & \xmark & \xmark & \cmark \\
             Default Parameter Values & \xmark & \xmark & \cmark \\
            \textbf{Upload infected node list} & \xmark & \xmark & \cmark \\
            Multiple simulations & \cmark & \xmark & \xmark \\
            Diffusion trends & \cmark & \cmark & \cmark \\
            Time-lapsed visual update & \xmark & \cmark & \cmark \\
            Dual-diffusion comparison  (inter-model) & \cmark & \xmark & \cmark \\
            \textbf{Dual-diffusion comparison (intra-model)} & \xmark & \xmark & \cmark \\
            \textbf{Dual-diffusion comparison (single view)} & \xmark & \xmark & \cmark \\ 
            \textbf{Ground-truth comparison} & \xmark & \xmark & \cmark \\
            Downloadable reports & \cmark & \xmark & \cmark \\
            \textbf{Download Iteration Data} & \xmark & \xmark & \cmark \\
            \textbf{Multi-User Support} & \cmark & \xmark & \cmark \\ 
            \textbf{\# Diffusion models supported} & $3$ & $15$ & $13$\\
            \textbf{\# Random graph type supported} &$1$ &$4$ &$1$\\
            \textbf{\# Custom graph input supported} & \xmark & \xmark &$6$ \\
             \hline
            \end{tabular}}
        \end{table}
\section{Evaluation}
As discussed in Section \ref{sec:related_work}, very few visual analytics tools are available for information diffusion. With a prime focus on web interface tools for diffusion visualization, we compare \framework\ with two other web interfaces built for the same purpose - \texttt{Epinet} and \texttt{NDlib-Viz}. \texttt{Epinet} is a tool built upon the EpiModel \cite{epimodel} package and hosted through Shiny, an R package. \texttt{NDlib-Viz} is a visualization module built on top of \texttt{NDlib} \cite{ndlib}. This section performs a three-way evaluation of \framework\, comparing feature, performance, and user interaction level. The comparative study helps us provide an in-depth analysis of the supported functionalities and the easy-of-use of those functionalities from an end-user perspective for both \framework\ and the competing web interfaces. 
\label{sec:eval}
\subsection{Feature Evaluation}
\label{sec:feat_eval}
It is evident from the exhaustive feature comparison in Table \ref{tab:feature_comp} that \texttt{NDlib-Viz} presents a bare minimal set of features, which are provided by \texttt{Epinet} and \framework\ as well. In addition, \framework\ supports a wider array of features in terms of interactive visualization, advanced network statistics and comparative diffusion analysis, and extendable back-end APIs. Further the completeness of the interfaces can be observed from the last $3$ rows of Table \ref{tab:feature_comp}. While both \texttt{NDlib-Viz} and \framework\ provide large number of diffusion models to choose from ($15$ and $13$ respectively), \texttt{Epinet} falls far behind in this ($3$ models supported). Additionally, our system supports loading custom graphs in $6$ different standard graph formats. One area where both \texttt{Epinet} and \framework\ fall behind is the support for different variants of random graphs. We hope, in future iterations, to add support for varying random graph types other than the currently-supported Erdos-Renyi model.

\subsection{Performance Evaluation}
\label{sec:perf_eval}
\subsubsection{Performance based on ease and feasibility of actions}
Based on the user requirement survey, it is apparent that users would prefer a system that is easy to set up and provides support for scripting and customization. Based on these requirements, we did a comparative performance evaluation on the number of steps it takes to run an end-to-end diffusion setup. As it can be observed from Table \ref{tab:perf_action_comp},
being a hosted service \texttt{Epinet} has the least setup requirement; meanwhile, both \texttt{NDlib-Viz} and \framework\ require steps to download and get the interface running. Double the number of steps required in setting up \texttt{NDlib-Viz} stems from its dependency on $2$ projects that need to be set up in order for the visualization interface to work. It should be noted that even though \texttt{Epinet} has no setup requirements, the system times out without a warning and checkpoints, leading to loss of information. Meanwhile, \texttt{NDlib} and \framework\ do not time out unless the user kills the application. Owing to the complexity of `Network Model Estimation', \texttt{Epinet} requires a maximum number of steps to load a random network. In terms of steps to set up the initial network, both \texttt{NDlib-Viz} and \framework\ are pretty easy to get started with. Additionally, for the same number of steps, \framework\ also supports the loading of custom graphs. On average, keeping the ability to set different diffusion model-wise parameters constant, we see that all the interfaces require the same number of steps to run the diffusion model ($4$ steps). However, our model requires the least number of steps ($11$ steps) in terms of an end-to-end setup. The set of end-end workflow should comparable to \texttt{NDlib-Viz} ($13$ steps) and visibly better than \texttt{Epinet} ($15$ steps). Our empirical analysis is further corroborated by a human evaluation comparing the three systems (Section \ref{sec:com_user_study}) rating our system high on ease of learning and usage.
 
\begin{table}[]
        \centering
            \caption{Performance comparison among three web-based diffusion visualization tools --- \texttt{Epinet}, \texttt{NDlib-Viz} and \framework, based on their ability to complete an action. \xmark\ indicates action not supported by the interface.}
            \label{tab:perf_action_comp}
            \scalebox{0.75}{
            \begin{tabular}{p{0.5\linewidth}|p{0.1\linewidth}|p{0.11\linewidth}|p{0.08\linewidth}}
            \hline
            \textbf{Action} & \textbf{\texttt{Epinet}} & \textbf{\texttt{NDlib-Viz}} & \textbf{\framework}\\
            \hline
            \# steps to setup the interface & $1$ & $6$ & $3$ \\
            \# steps to load a random network  & $9$ & $3$ & $3$ \\
            \# steps to load a custom network  & \xmark & \xmark & $3$ \\
            \# steps to run a standard diffusion model & $4$ & $4$ & $4$ \\
            \# steps to run a custom diffusion script & \xmark & \xmark & $5$ \\
            \# steps to save/download diffusion results & $2$ & \xmark & $2$ \\
            \# steps to run an end-to-end SI model for $100$ iterations on random graph with $200$ nodes & $15$ & $13$ & $11$\\ \hline
             \end{tabular}}
        \end{table}

\begin{table}[]
    \caption{Performance comparison among three web-based diffusion visualization tools --- \texttt{Epinet}, \texttt{NDlib-Viz} and \framework, based on their ability to load and run diffusion models for varying graph sizes. \xmark\ indicates action not supported by the interface. The time to load a random network and run a diffusion model is recorded in seconds as an average of $5$ runs. All the systems are evaluated on locally-hosted Chrome browser (Version 101.0.4951.64) on ThinkPad E480 with $16$GB RAM.}
    \label{tab:perf_custom_comp}
    \scalebox{0.75}{
    \begin{tabular}{c|c|c|c|c |c|c}
    \hline
    \multirow{2}{*}{\textbf{\# Nodes}} & \multicolumn{2}{c|}{\textbf{\texttt{Epinet}}} & \multicolumn{2}{c|}{\textbf{\texttt{NDlib-Viz}}} & \multicolumn{2}{c}{\textbf{\framework}} \\
    & \multicolumn{1}{c|}{\textbf{Load network}} & \multicolumn{1}{c|}{Run Diffusion} & \multicolumn{1}{c|}{\textbf{Load network}} & \multicolumn{1}{c|}{\textbf{Run Diffusion}} & \multicolumn{1}{c|}{\textbf{Load network}} & \multicolumn{1}{c}{\textbf{Run Diffusion}} \\  \cline{1-7}
    $100$  & $1.24$ & $1.36$ & $0.64$  & $1.13$ & $2.28$ & $0.982$ \\
    $300$  & $1.40$ & $1.49$ & $0.694$ & $1.17$ & $2.00$ & $1.16$  \\
    $500$  & $1.39$ & $2.00$ & $1.13$  & $1.08$ & $1.96$ & $1.19$  \\
    $700$  & $1.23$ & $2.18$ & $1.37$  & $1.44$ & $2.14$ & $1.28$  \\
    $1000$ & $1.30$ & $2.47$ & $2.06$  & $2.08$ & $2.69$ & $1.42$  \\
    $3000$ & $1.94$ & $7.47$ & $14.24$ & $6.84$ & $6.72$ & $3.96$  \\
    $6000$ & $6.62$ & $39.02$ & \xmark & \xmark & $14.04$ & $9.07$ \\
    $10000$ & \xmark & \xmark & \xmark & \xmark & $14.57$ & $13.66$ \\ \hline
    \end{tabular}
    }
\end{table}

\begin{figure}[]
    \centering
    \subfloat[]{\includegraphics[width=0.475\textwidth]{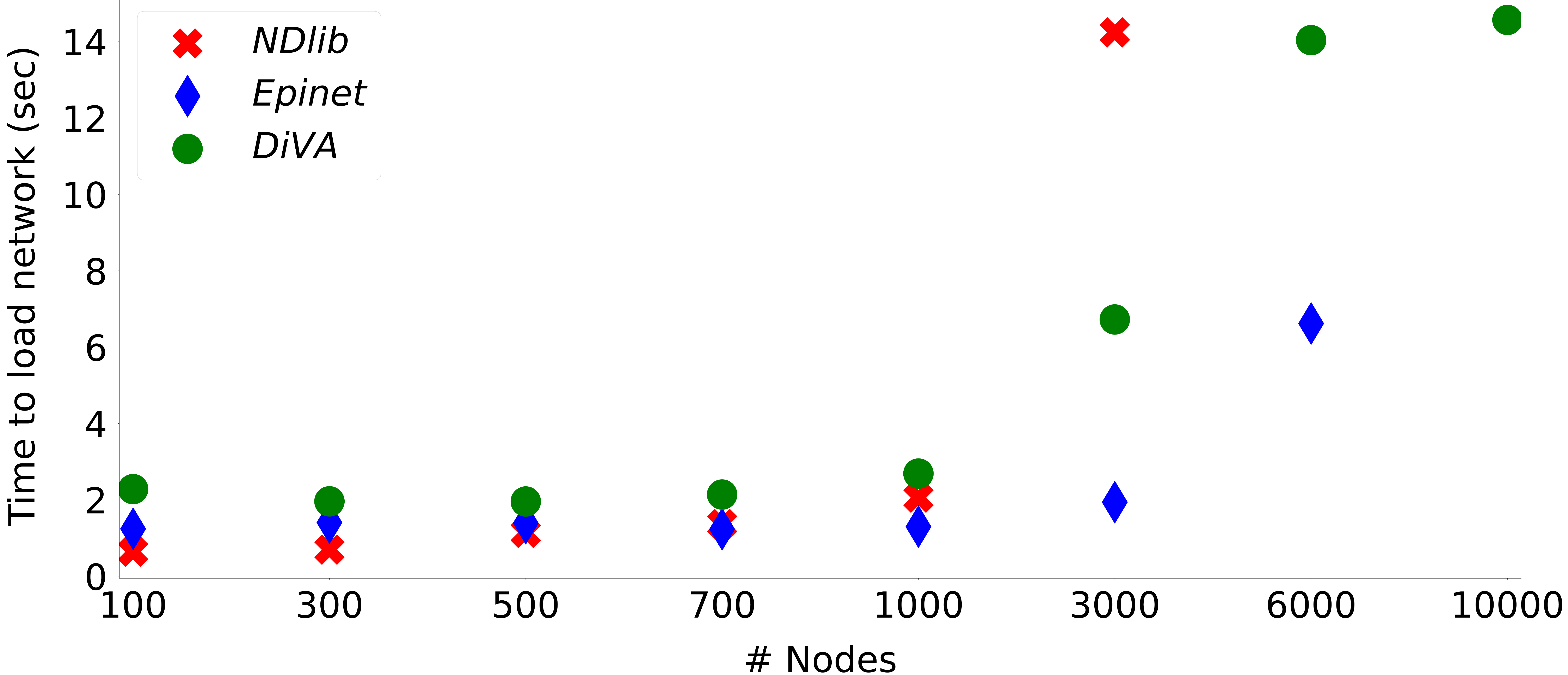}}\hspace{1mm}
    \subfloat[]{\includegraphics[width=0.475\textwidth]{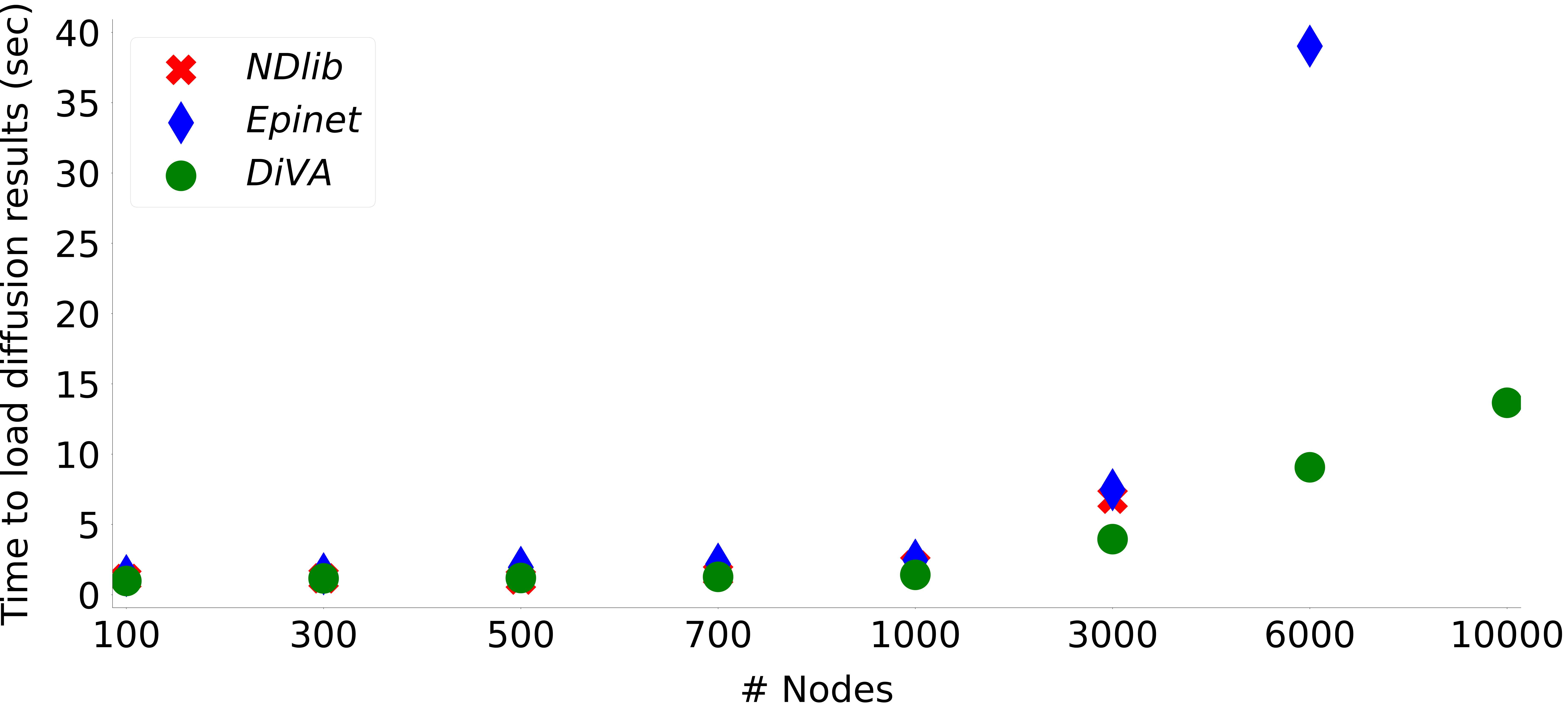}}
    \caption{Plots accompanying Table \ref{tab:perf_custom_comp}. The performance is measured by the time taken to perform: a) Time to load a network with a given number of nodes. b) Time to run the diffusion simulation for $100$ steps with the network loaded in part (a).}
    \label{fig:perf_comp}
\end{figure}
 
\subsubsection{Performance based on scalability}
One can observe from Table \ref{tab:perf_custom_comp} and subsequent plots in Figure \ref{fig:perf_comp} that under similar system configuration, \texttt{Epinet} and \texttt{NDlib-Viz} became unresponsive once the number of input nodes exceeds $10,000$ and $6000$, respectively. For shorter, the slightly higher time for \framework\ to load the initial network can be attributed to the time it takes to the Force-Atlas algorithm that generates the layout before the network eventually gets displayed. Meanwhile, under the same diffusion settings, we observe that the time to obtain the results for the SI diffusion model (used in this evaluation) is much faster for \framework in comparison to  \texttt{Epinet} and \texttt{NDlib-Viz}. At $3k$ nodes (the largest network size available to compare all three systems), we observe that \framework\ produces a small gain of $2.88$ and $3.51$ seconds over \texttt{NDlib-Viz} and \texttt{Epinet}, respectively.  Epimodel\footnote{\url{https://rdrr.io/github/statnet/EpiModel/}} is the underlying diffusion modeling package used by \texttt{Epinet}. Meanwhile, both \texttt{NDlib-Viz} and \framework\ rely on NetworkX and NDlib to perform the underlying diffusion modeling. The difference in the performance of the three systems can, in part, be attributed to how the underlying packages handle diffusion simulations. Since both \texttt{NDlib-Viz} and \framework\ rely on NetworkX and NDlib, we believe the performance boost in \framework\ is due to (a) superior graphing library in three.js, and (b) the way we stream response from backend to frontend. 
In addition to becoming unresponsive on large networks, the technologies used for \texttt{Epinet} and \texttt{NDlib-Viz} restrict performance on smaller networks as well. For example, while \texttt{NDlib-Viz} uses D3.js and D3-force to render the network, it uses a Canvas component which limits the size of the network and causes a considerable drop in the frame rate while zooming and panning the network. Although \framework\ also uses D3-force to set the network's layout, it does not suffer from these restraints. It can be attributed to final visualizations plotted on the WebGL component using Three.js. It makes \framework's network visualization relatively interactive and allows a high frame rate on large graphs, as evident from Table \ref{tab:diva_perf}. Further, the scalability of \framework\ in terms of space and time is again highlighted in Table \ref{tab:diva_perf}. Our system can load and operate on networks ranging from as small as $400$ nodes and $1.5k$ edges to networks as large as $68k$ nodes and $4.5M$ edges. The largest network we tested on ($68k$ nodes and $4.5M$ edges) took $\approx 6$ mins to load. If this same network is saved as .diva and loaded again, the load time reduces to $2.3$ mins. Not only is \framework\ scalable in terms of time, but also space. An edge set of $4.5M$ takes up less than $1kB$ of RAM.
\begin{table}[]
    \centering
        \caption{Performance of \framework\ w.r.t different system parameters. The simulations were run on Google Chrome version 86.0.4240.75 on a system with Intel Core i7-9750H processor, an Nvidia GTX 1650 4GB GDDR5 Graphics Processing Unit, 1TB PCIe NVMe SSD and a 60fps Full-HD monitor.}
        \label{tab:diva_perf}
        \scalebox{0.75}{
        \begin{tabular}{c|c|c|c|c|c}
            \hline
            \bfseries{\# Nodes} & \bfseries{\# Edges} & \bfseries{Frame rate} & \bfseries{RAM used} & \bfseries{Time to load} & \bfseries{Time to load}\\
            & & & & & \bfseries{from .\texttt{diva} file}\\
            \hline
            400 & 1500 & 60 fps & 10 MB & 5 sec & 1 sec\\
            1k & 3800 & 60 fps & 13 MB & 6 sec & 1 sec\\
            1.5k & 5100 & 60 fps & 14 MB & 6 sec & 1 sec\\
            10k & 100k & 60 fps & 38 MB & $\approx$ 25 sec & 6 sec\\
            5k & 150k & 60 fps & 73 MB & $\approx$ 23 sec & 8 sec \\
            25k & 700k & 60 fps & 159 MB & $\approx$ 1.7 min & $\approx$ 35 sec\\
            35k & 1.33M & 60 fps & 538 MB &  $\approx$ 3 min & $\approx$ 1.1 min\\
            \textbf{68k} & \textbf{4.5M} & \textbf{35 fps} & \textbf{925 MB} & \textbf{$\approx$ 6 min} & \textbf{$\approx$ 2.3 min} \\
            \hline
        \end{tabular}}
\end{table}
\subsection{User Evaluation}
\label{sec:user_eval}
We conducted user studies that spanned the development of the tool. In the first study, the participants got a chance to review \framework\ and the two other competitor web interface tools. In the second stage of the user study, \framework\ was exclusively tested for its accessibility and usability. 
    
\subsubsection{Comparative User Study}
\label{sec:com_user_study}
As a part of our first user study, we reviewed the competing web-based diffusion visualization tools (\texttt{Epinet}, \texttt{NDlib-Viz} and \framework). The participants (human subjects) reviewed all the services hosted for them and provided comparative feedback on the three systems. The following steps were considered in  order to reduce bias towards any single tool: 
\begin{itemize}
    \item As our tool is a web-based system, we perform comparative analysis with two other established diffusion visualization and analytic tools that are also web interfaces by themselves. This makes sure the user experience is uniform.
    \item While \texttt{Epinet} is already hosted, our framework and \texttt{NDlib-Viz} need to be locally set up first. Thus, we hosted both the services and let the users access all three services from an anonymous server. Additionally, using a hosted service further unifies the user experience and focuses on the tools, not the installation and dependencies. 
    \item The human subjects were not initially told that \framework\ is a new tool; it was instead posed as another established tool.
\end{itemize}
The study inducted $44$ human evaluators (27/17 males/females). The age of the human subjects ranged from $24$-$40$ years. Each participant took $\sim 30$ minutes to fill the entire survey. At the beginning of the survey, we asked participants about their familiarity with information diffusion in general (`Yes,' `No,' `Somewhat'). During the survey analysis, based on the above answer, we divided the participants into three groups --- experts,  non-experts, and enthusiasts, respectively. Out of the $44$ participants, we had $11$ experts, $14$ non-experts and $19$ enthusiasts.

While \framework\ supports custom networks, \texttt{Epinet} and \texttt{NDlib-Viz} only provide support to set up a random network; so to begin with, all the subjects were asked to set a random network consisting of $1.5$k nodes and $5$k edges. Though small, these numbers ensured the competitive services were able to load the network smoothly (Section \ref{sec:perf_eval}). Once the network was established, the subjects were asked to run the SI model of diffusion (for non-expert users, we provided a link to the model's documentation\footnote{https://ndlib.readthedocs.io/en/latest/reference/models/epidemics/SIm.html}) and then to evaluate the reports generated for that diffusion. Once the participants completed the workflow, they comparatively rated the three interfaces on a scale of $1$-$3$ ($3$ being the best rating). The ratings concerned (a) \textit{easy of use and learning}, (b) \textit{minimal actions to achieving analysis}, and (c) \textit{overall system capabilities}.  A summary of these ratings is provided in Table \ref{tab:survey_results}. While \framework\ is given the highest rating in each category, when it comes to \textit{ease of learning}, our tool and \texttt{NDlib-Viz} are seemly closers in rating. This puts \framework\ in a favorable position as our framework is more feature-rich yet easy to follow. Notably, it is in providing the \textit{overall system capabilities} that our tool outshines the others by a considerable margin. To reduce the barrier to entry in information diffusion, we strike a challenging balance between providing a feature-rich yet easy-to-use tool. Our survey results point out that while we still have a long way to go, \ the framework\ is a step in the right direction.

    \begin{table}[!h]
        \caption{Comparative rating of the interfaces --- \texttt{Epinet}, \texttt{NDlib-Viz} and \framework\ as perceived by the participants. The results are summarised as \% of total user ($44$) agreeing upon a  rating for the respective tool. Rating scales from $3$ (best) to $1$ (worst). Results are rounded off to one decimal.}
        \label{tab:survey_results}
        \renewcommand{\arraystretch}{0.8}
        \centering
        \scalebox{0.75}{
        \begin{tabular}{p{0.25\linewidth}|c|c|c|c|c}
        \hline \textbf{Feature} & \textbf{Rating} & \textbf{\texttt{Epinet}} & \textbf{\texttt{NDlib-Viz}} & \textbf{\framework}\\
        \hline
        \multirow{3}{\linewidth}{Ease of learning} & $\textbf{3}$ & $13.6$ & $40.9$ & $\textbf{43.2}$\\
        & $2$ & $47.7$ & $43.2$ & $38.6$\\
        & $1$ & $38.6$ & $15.9$ & $18.2$\\
        \hline
        
        \hline
        \multirow{3}{\linewidth}{Ease of use} & $\textbf{3}$ & $13.6$ & $40.9$ & $\textbf{52.3}$\\
        & $2$ & $38.6$ & $43.2$ & $38.6$\\
        & $1$ & $47.7$ & $15.9$ & $9.1$\\
        \hline
        
        \hline
        \multirow{3}{\linewidth}{Minimal actions for analysis} & $\textbf{3}$ & $18.2$ & $38.6$ & $\textbf{52.3}$\\
        & $2$ & $47.7$ & $40.9$ & $38.6$\\
        & $1$ & $34.1$ & $20.5$ & $9.1$\\
        \hline
        
        \hline
        \multirow{3}{\linewidth}{Overall system capabilities} & $\textbf{3}$ & $27.3$ & $11.4$ & $\textbf{63.6}$\\
        & $2$ & $54.5$ & $50.0$ & $27.3$\\
        & $1$ & $18.2$ & $38.6$ & $9.1$\\
        \hline
        \end{tabular}}
    \end{table}
    
\subsubsection{Accessibility Study}
For our second user study, we approached $58$ users\footnote{Consents and approvals were taken from all the users that participated in our study.} ($39$ male and $19$ female), where most of the users belonged to the age group of $24$ to $40$. While some of the human subjects had already participated in our Comparative User Study (Section \ref{sec:com_user_study}), we also inducted new evaluators. It helps reduce the familiarity bias\footnote{\url{https://en.wikipedia.org/wiki/Familiarity_heuristic}} of having interacted with \framework\ before. At the same time, we did not replace all the previous evaluators to maintain the continuity of a long-range study for future user evaluations of the tool. We again asked the users about their familiarity with the concepts of diffusion modeling. $18$ were marked as ``expert" users --- who had prior experience with network diffusion tools. The $40$ ``novice" users did not have any experience with a similar tool. However, they explicitly mentioned that they were interested in information diffusion. As a metric of evaluating usability, we employed the System Usability Scale (SUS) metric \cite{SUS,SUSSurvey} (see Appendix \ref{app:sus_ap} for more details). Based on a ten-item questionnaire, the SUS score assigns a numerical value between $0$ and $100$ to the system under review. A score above 68 is considered to be average\footnote{\url{https://www.usability.gov/how-to-and-tools/methods/system-usability-scale.html}}.
    
    \begin{table*}[hbt!]
    \caption{Statistics of user responses for SUS evaluation. The values indicate the number of users who marked that option of the corresponding question. ("S. Disagree": Strongly Disagree, "S. Agree": Strongly Agree).}
    \centering
    \scalebox{0.75}{
    \begin{tabular}{p{0.48\linewidth}|c|c|c|c|c}
    \hline
    \textbf{Question} & \textbf{S. Disagree} & \textbf{Disagree} & \textbf{Neutral} & \textbf{Agree} & \textbf{S. Agree} \\ \hline
    I think I would like to use this system frequently. & 1 & 3 & 8 & 37 & 9  \\ \hdashline 
    I found the system unnecessarily complex. & 21 & 25 & 9 & 3 & 0 \\   \hdashline 
    I thought the system was easy to use. & 0 & 3 & 7 & 24 & 24  \\   \hdashline 
    I think that I would need the support of a technical person to be able to use this system. & 17 & 23 & 10 & 6 & 2  \\   \hdashline 
    I found the various functions in this system were well integrated. & 0 & 2 & 1 & 29 & 26 \\   \hdashline 
    I thought there was too much inconsistency in this system. & 33 & 19 & 4 & 2 & 0 \\   \hdashline 
    I would imagine that most people would learn to use this system very quickly. & 0 & 5 & 6 & 30 & 17  \\  \hdashline 
    I found the system very cumbersome to use. & 33 & 16 & 8 & 1 & 0 \\   \hdashline 
    I felt very confident using the system. & 0 & 2 & 9 & 30 & 17  \\   \hdashline 
    I needed to learn a lot of things before I could get going with this system. & 11 & 20 & 18 & 4 & 5 \\  \hline
    \end{tabular}}
    \label{tab:sus-metrics}
    \end{table*}
    
The study proceeded as follows:
    \begin{itemize}
        \item Novice users were given a brief explanation about networks and information diffusion with the help of some epidemic models.
        \item The users were tasked with analyzing the spread of a virus over a dummy network with $1k$ nodes.
        \item The users were given a set of instructions to run a diffusion algorithm along with tinkering with the customization and analytical features of \framework\, enabling them to explore both the tool's technical and user experience aspects.
        \item The users were also asked to complete the same task by independently exploring the system and algorithm configurations.
        \item Finally, the users were asked to fill out the survey for us to calculate the SUS score.
    \end{itemize}
    
Table \ref{tab:sus-metrics} lists the questions and summarizes the user response. Note that the set of questionnaires used was the standard SUS survey template, and we did not tweak any part of it. It helps reduce any bias that can be introduced from our side. Additionally, before the survey, we deployed the tool and shared the link of the hosted application that the users accessed via their browsers. Hence creating a more consistent experience for the users and further mitigating the user bias.

Table \ref{tab:sus_stats_table} lists different statistical measures calculated on the SUS scores. The Upper Bound of the SUS score was $97.5$ for all user types. From our evaluation, we obtained a mean score of $77.2$ (among all users), which is quite above the average score of 68, rating \framework\ as ``Acceptable" (according to the SUS guideline). Individually for the expert and novice users, their mean scores were way above the threshold ($83.1$ and $74.6$ respectively). As expected, the expert users had a higher mean SUS given their familiarity with diffusion terminology; they could adapt faster to the interface. Even though many of our participants were a novice, they all felt confident in using the system by themselves. They did not seem to need additional technical assistance in exploring the system. This validates the design decisions made regarding where different tool components are placed.
    
    \begin{table}[H]
    \centering
        \caption{System Usability Scale (SUS) statistics of our user study.}
        \label{tab:sus_stats_table}
        \scalebox{0.75}{
        \begin{tabular}{m{0.2\textwidth}|m{0.1\textwidth}|m{0.1\textwidth}|m{0.1\textwidth}|m{0.1\textwidth}|m{0.12\textwidth}|m{0.175\textwidth}}
                    \hline
                    \textbf{User Type} & \textbf{Count}& \textbf{Mean} & \textbf{S.D} & \textbf{Median} & \textbf{Lower Bound} & \textbf{$75^{th}$ percentile} \\
                    \hline
                    \textbf{Novice Users} & 40 & 74.6 & 12.7 & 76.25 & 40 & 82.5\\ 
                    \textbf{Expert Users} & 18 & 83.1 & 9.79 & 85 & 60 & 89.3\\ 
                    \textbf{All Users} & 58 & 77.24 & 12.46 & 77.5 & 40 & 86.88\\ 
                    \hline
        \end{tabular}}
    \end{table}

\begin{figure}[!ht]
    \centering
    \includegraphics[width=\textwidth]{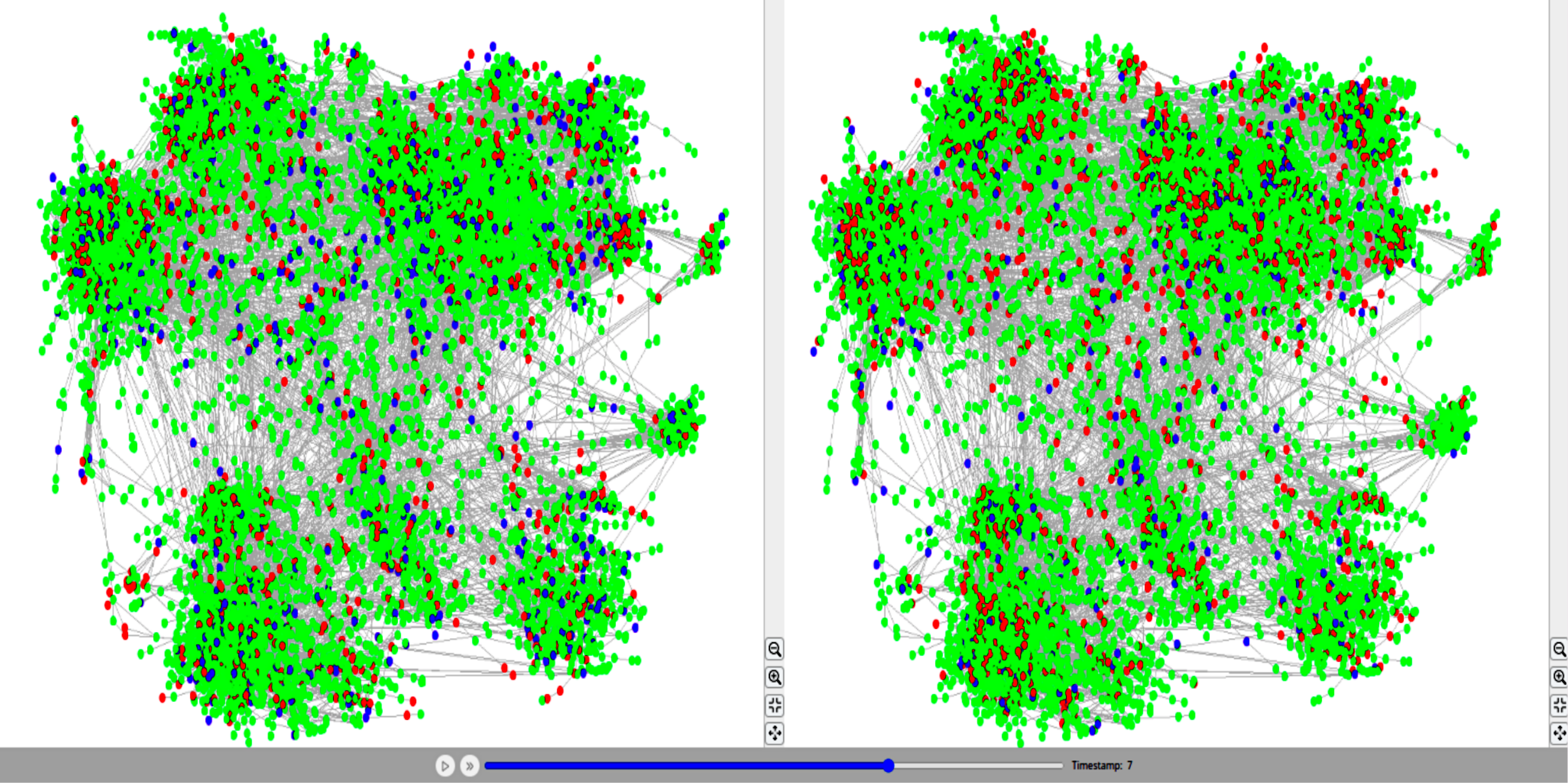}
    \caption{Results of our simulations on the FEATHER network. These observations are for $t=7$ in the \texttt{Split View} of the dual visualization mode. The green, red and blue nodes represent the susceptible, infected and recovered nodes, respectively. The recovery rate for the SIR model on left is $\gamma=0.1$. Meanwhile, keeping other factors constant, the recovery rate for the SIR model on the right is $\gamma=0.05$.}
    \label{fig:case_study_iter}
\end{figure}

\section{Case Study}
\label{sec:case_study}
In this section, we consider a standard dataset, called LastFM-Asia of social network users, first introduced in the FEATHER network analysis \cite{FEATHER}. We run some diffusion simulations on this.

\textbf{Dataset Description:} The dataset was curated from LastFM public API (March 2020) for the Asian region consisting of users with $18$ country labels. There are $7624$ user nodes, which are connected via $27806$ mutual friendship links. The original research \cite{FEATHER} aimed to classify users into their country of origin based on their friendship network and node-level features of artist affiliation.

In our setup, we use only the friendship network of LastFM-Asia and employ the dataset for diffusion prediction instead, assuming that an Internet virus spreads across the network and a user, based on interaction with his/her friends, exposes more users to the virus.

\textbf{Aim.} For a case of an Internet virus that is spreading on the LastFM-Asia network, how many systems can we recover with varying rates of recovery?

\textbf{Diffusion Algorithm.} In order to obtain recovered nodes in the network, the closest diffusion model would be the SIR (susceptible-infected-recovered) model. SIR model has two parameters -- (i) the rate of infection ($\beta$), which in our case is the rate at which the computer virus is affecting the users, and (ii) the rate of recovery ($\gamma$), which in our case is the rate at which users can recover their systems.

\textbf{Simulation Environment.} All the simulations are run on a locally-hosted Chrome browser (Version 101.0.491.64) on ThinkPad E480 with 16GB RAM.

\textbf{Steps for simulation.}
\begin{itemize}
    \item Bring up an instance of \framework, and load the LastFM-Asia network.
    \item Switch to \texttt{Compare View} tab.
    \item Select SIR as the first diffusion algorithm with parameters $\beta_1=0.05$, $\gamma_1=0.1$.
    \item Select SIR as the first diffusion algorithm with parameters $\beta_2=0.05$, $\gamma_2=0.05$.
    \item For our use simulation we set the \texttt{Maximum Iterations} $=10$, \texttt{Fraction Infected} $=0.1$.
    \item Run the iteration, and visualize the results with different $\gamma$ rates.
\end{itemize}

\begin{figure}[!h]
    \centering
    \includegraphics[width=0.5\textwidth]{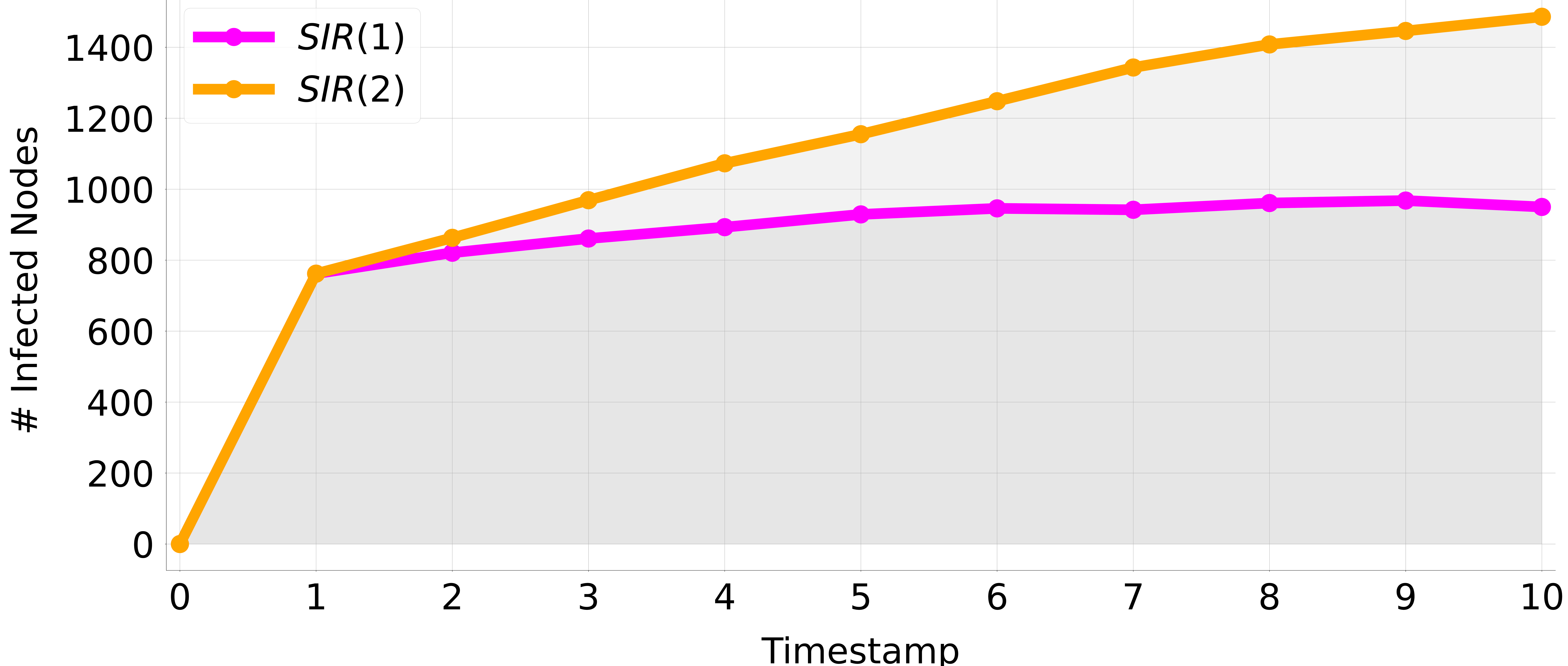}
    \caption{The number of infected nodes for the two simulations as obtained from the \texttt{Report View Tab}. SIR (1) and SIR (2) refer to models with recovery rates $0.1$ and $0.05$ respectively.}
    \label{fig:case_study}
\end{figure}

\textbf{Evaluating the Results.}
\begin{itemize}
\item Firstly, we obtain the network statistics as reported in the original paper, and as evident from Table \ref{tab:case_study}, we can reproduce the same.
\item Secondly, from Table \ref{tab:case_study} we notice that keeping all others same, a higher rate of recovery translated to more nodes saved and restored from the computer virus. At the end of the simulation, we will be able to recover almost $1.5x$ the number of nodes ($806$ vs. $457$) for a higher recovery rate. The same can be observed from the difference in the number of infected vs. recovered nodes from the infection plots in Figure \ref{fig:case_study}. At timestamp $t=2$ the delta increase in number of infected nodes is $42$, while at end of the simulation ($t=10$) the delta increase in number of infected nodes is as high as $536$.   
\item For a SIR model with green, blue, and red nodes specifying the susceptible, infected, and
recovered nodes, respectively, we can visually observe the impact of the high recovery rate in the first SIR model by the higher number of blue nodes in the left split of Figure \ref{fig:case_study_iter}. Complementary information for the lower rate of recovery in the second SIR model can be visually observed by a lower number of blue nodes in the left split of Figure \ref{fig:case_study_iter}. The right split subsequently has a higher volume of red nodes.
\end{itemize}

\begin{table}[!h]
\centering
\caption{Evaluating the results for simulation: (a) Reporting the reproduced network statistics. (b) listing out the number of recovered nodes in the system for different rate of recoveries ($\gamma$).} 
\label{tab:case_study}    
\begin{tabular}{c|c|c}
\hline
\textbf{Statistics}   & \textbf{Original} & \textbf{Reproduced} \\ \hline
\# Nodes     & $7624$    & $7624$      \\
\# Edges     & $27806$    & $27806$      \\
Density      & $0.0009$   & $0.0009$     \\
Transitivity & $0.179$    & $0.179$  \\ \hline   
\end{tabular}
\begin{tabular}{c|c|c}
\hline
\multirow{2}{*}{\textbf{Timestamp}} & \multicolumn{2}{c}{\textbf{\# Recovered Nodes}}\\ \cline{2-3}
& \textbf{$\gamma=0.1$} & \textbf{$\gamma=0.05$} \\\hline
$0$         & $0$                         & $0$                          \\
$1$         & $0$                        & $0$                         \\
$2$         & $73$                       & $21$                         \\
$3$         & $153$                       & $57$                        \\
$4$         & $238$                       & $95$                        \\
$5$         & $326$                       & $154$                        \\
$6$         & $426$                       & $207$                       \\
$7$         & $533$                       & $260$                        \\
$8$         & $624$                       & $324$                        \\ 
$9$         & $710$                       & $389$  \\
$10$         & $806$                       & $457$
\\ \hline     
\end{tabular}
\end{table}

\section{Conclusion and Future Work}
\label{sec:conclusion}
Interest in information diffusion has increased with the increasing availability of large-scale networks. We present \framework, an interactive web-interface-based visualization tool that scales well for large networks. It provides three features ---  support for large networks (along with the ability to save their state for future use), simulation and visualization of diffusion algorithms, and comparison of two diffusion algorithms in a split window. Additionally, \framework\ allows high customizability for the users in terms of the layout and visuals of the network, provides diffusion analysis reports, and facilitates the users to visualize a ground-truth diffusion and even compare it to a simulation of another algorithm. Despite the large array of features supported by the system, we believe that the systematic workflow makes it easy for users of all experience levels to use the tool effectively. 

The current version of \framework\ uses NetworkX for network modeling, given its popularity, community support, and pure Python implementation. We plan to provide integration for igraph in further increments. We aim to add some more advanced diffusion algorithms. 
The world is ever-changing, and so are the networks. We hope to incorporate more tools and feature sets such as introducing spatial information, geotagging, user influence analysis, and deeper community analysis into \framework\ to make it more useful for both expert and novice users working in network science. Additionally, to incorporate the temporal nature of networks, we hope to extend \framework\ for dynamic networks and compare two different networks (network-to-network comparison of diffusion). Within the existing setup, the underlying network diffusion library (NDlib \cite{ndlib}) has limited support for snapshot and interaction diffusion executions. Only $6$ diffusion algorithms are available in dynamic mode. We hope to leverage this setup to introduce support for dynamic networks in the future.   

\section*{Acknowledgement}
We would like to thank the support of Prime Minister Doctoral Fellowship (SERB India), Ramanujan Fellowship (SERB, India), and the Wipro research grant. We also thank all the members of the LCS2 Lab and other human subjects for participating in the tool evaluation process and providing us with constructive feedback. 

\bibliography{diva.bib}

\appendix
\section{Uploading Custom Diffusion Algorithm}
\label{app:Custom_algo_code}
The users can upload their algorithm in the form of a python script. The iterations that form the output of the template code should form a list of python dictionaries, where each dictionary corresponds to an iteration and indicates the change in status of each node after each iteration. The dictionary only needs to specify the states of those nodes that have undergone a state change since the previous iteration. Furthermore, the dictionaries must also report the total number of nodes in each state at each iteration. An example code and output is shown below (Code \ref{fig:custom_iter}).
  
{\tiny\label{fig:custom_iter}
\begin{lstlisting}[language=python,firstnumber=1, basicstyle=\small]
[ 
    {"iteration": 0, 
     "status": {"A":1,"B":0,"C":0,"D":0},
     "node_count": {"0":4,"1":1,"2":0}},
    {"iteration": 1, 
     "status": {"A": 2,"B":1,"D":1}, 
     "node_count": {"0":1,"1":3,"2":1}} 
]
\end{lstlisting}
}
{\small \label{fig:custom}
\begin{minted}{python}
import networkx as nx
import json
import pandas as pd
import app
# Other imports needed ...

class Model():
    def __init__(self, G, seeds=None, fraction_infected=None, iterations=10):
        """ 
            G -> NetworkX graph.
            seeds -> An array of actual labels of the seed nodes
            fraction_infected -> float[0,1]
            iterations -> Max iterations for a single simulation
            Returns -> null
        """
    def run_model(self):
        """
            Return the iterations in the same format as NDlib does. 
            Returns -> List of Dictionary
        """
    ### Other functions needed ...
\end{minted}
}

\section{System Usability Scale (SUS)}
\label{app:sus_ap}
The System Usability Scale was a metric created by John Brooke in $1996$ as a "quick and dirty" way to measure the usability of products \cite{SUS} and has been used extensively to test and evaluate numerous systems and applications. The scale consists of a ten-item questionnaire:
\begin{enumerate}
    \item I think I would like to use this system frequently.
    \item I found the system unnecessarily complex.
    \item I thought the system was easy to use.
    \item I think that I would need the support of a technical person to be able to use this system.
    \item I found the various functions in this system were well integrated.
    \item I thought there was too much inconsistency in this system.
    \item I would imagine that most people would learn to use this system quickly.
    \item I found the system very cumbersome to use.
    \item I felt very confident using the system.
    \item I needed to learn many things before going on with this system.
\end{enumerate}

The SUS score presents users with five options for each question, where "Strongly Disagree" has a score of $1$ and "Strongly Agree" has a score of $5$, and the others are uniformly distributed between these two. Once the user fills up the SUS score form, the score is calculated as follows: 
\begin{enumerate}
    \item For every odd-numbered question, subtract $1$ from the score.
    \item For every even-numbered question, subtract the score from $5$.
    \item Sum up the new values and multiply them by $2.5$.
    \item The final value is between $0$and $100$ and gives us the SUS score for one user.
\end{enumerate}

Here, the response towards the "Strongly Agree" area is considered suitable for the odd-numbered statements (meant to be positive for the system). In contrast, this response is considered negative for the even-numbered statements. As per Bangor et al. \cite{SUSSurvey}, a SUS score of 68 is considered an average score. 

\section{Input Graph Formats Supported by \framework}
\label{app:input_graph}
\begin{figure}[!h]
    \centering
    \subfloat[]{\includegraphics[width=0.3\textwidth]{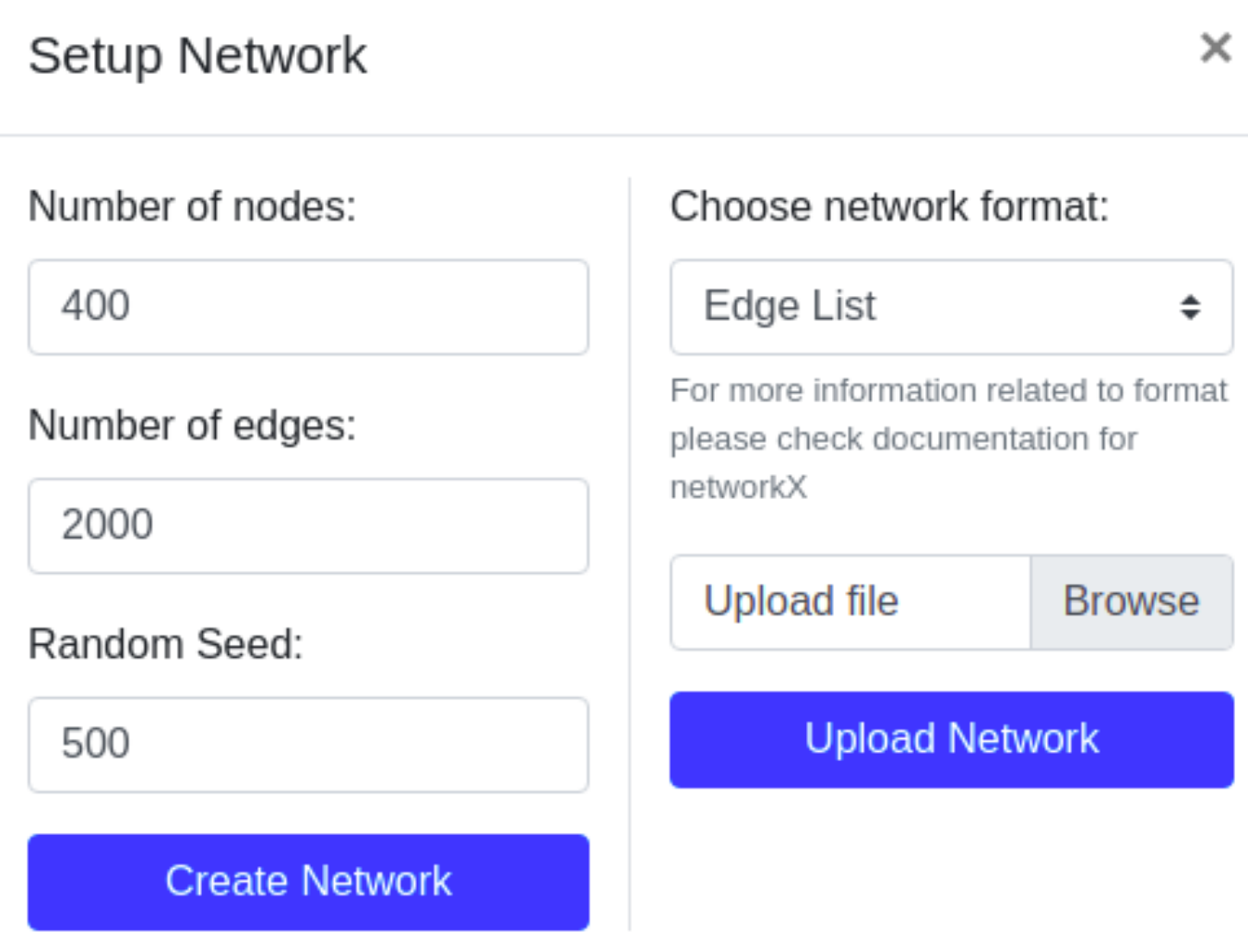}}\hspace{5mm}
    \subfloat[]{\includegraphics[width=0.3\textwidth]{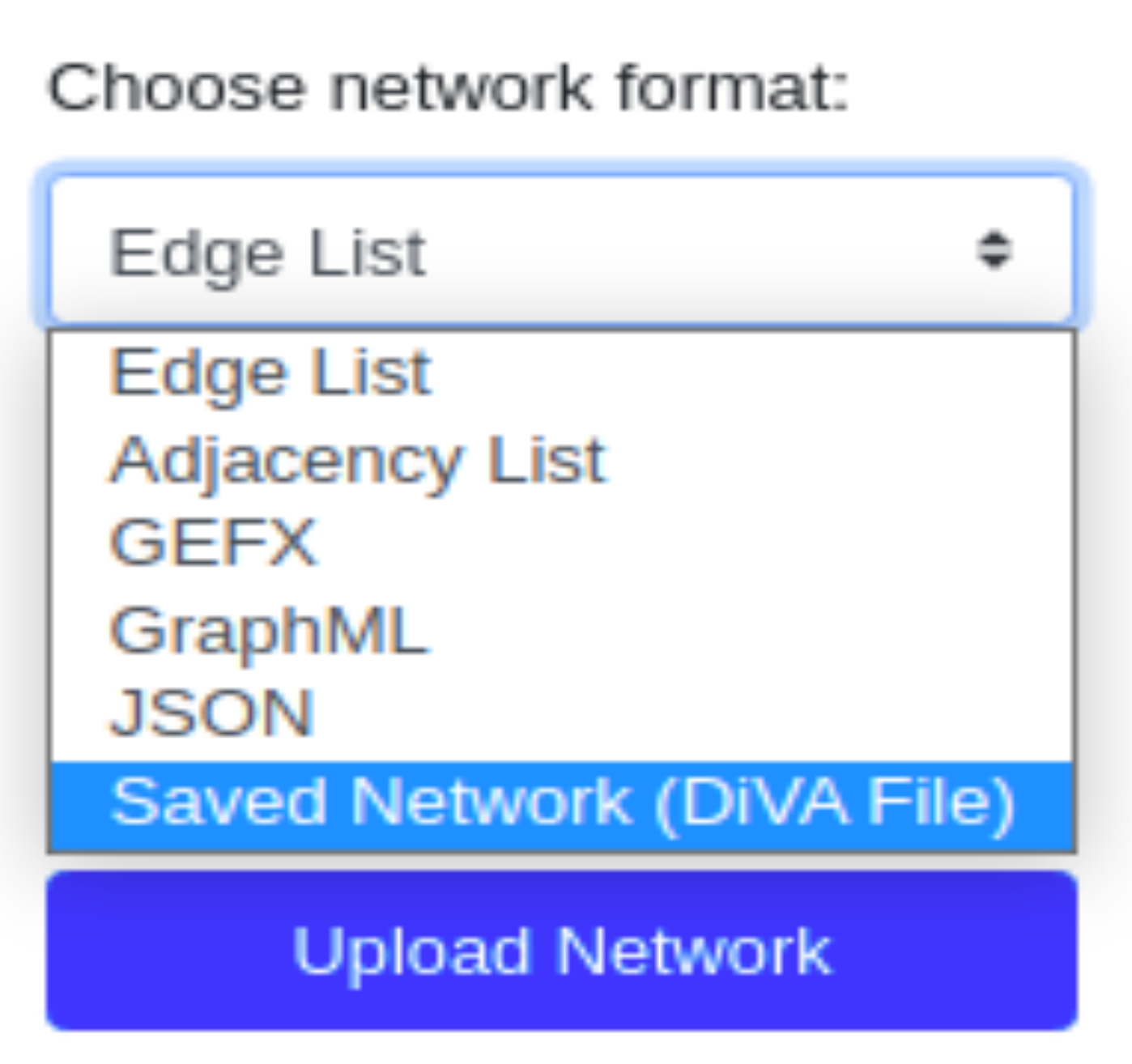}}
    \caption{teps for setting up the initial network --- (a) Users can select either a random graph or upload a custom one. (b) The network formats supported by \framework.}
    \label{fig:diva_net_set}
\end{figure}

Apart from the ER random graph model, we also provide support for uploading custom graphs in the following NetworkX supported formats -- Edgelist, Adjacency list, GEFX, JSON, and Graph ML. Various input formats as visible on the interface are depicted in Figure \ref{fig:diva_net_set}.

\section{Diffusion Models Supported by \framework}
\label{app:EP}
As it is difficult to accommodate the wide variety of information diffusion models available, in the current version of \framework, we provide the list of models as supported in the back-end by \texttt{NDlib} in Table \ref{tab:ep_models}. For details regarding these diffusion models and their limitations, readers are referred to these surveys of diffusion models \cite{Li2017ASO, Britton2010StochasticEM}.
    \begin{table}[]
        \caption{Diffusion models supported by \framework.}
        \label{tab:ep_models}
                \scalebox{0.75}
                {\begin{tabular}{p{0.28\textwidth}|p{0.3\textwidth}|p{0.38\textwidth}}
                    \hline
                    \textbf{Model Name} & \textbf{Input Parameters} & \textbf{Output Class Codes}\\
                    \hline
                    SI & beta & 0: susceptible, 1: infected\\ \hdashline
                    SIR & beta, gamma & 0:susceptible, 1: infected, 2: removed\\ \hdashline
                    SIS & beta, lambda & 0: susceptible, 1: infected\\ \hdashline
                    SEIS & alpha, beta, lambda & 0: susceptible, 1:infected, 2: exposed\\ \hdashline
                    SEIR & alpha, beta, gamma & 0: susceptible, 1: infected, 2: exposed, 3: removed\\ \hdashline
                    Threshold & node threshold & 0: susceptible, 1: infected\\ \hdashline
                    Generalised Threshold & tau, mu, node threshold & 0: susceptible, 1: infected\\ \hdashline
                    Profile & blocked, adopter rate, node profile & -1: blocked, 0: susceptible, 1: infected\\ \hdashline
                    Profile Threshold & blocked, adopter rate, node profile, node threshold & -1: blocked, 0: susceptible, 1: infected\\ \hdashline
                    Kertesz Threshold & adopter rate, Percent blocked, node threshold & -1: blocked, 0: susceptible, 1: infected\\ \hdashline
                    Independent Cascades & edge threshold & 0: susceptible, 1: infected, 2: removed\\ \hdashline
                    User Defined & User Defined & User Defined\\ \hdashline
                    Ground Truth & N/A & User Defined\\ \hline
                \end{tabular}}
            \end{table}
            
\section{Data View}
\label{app:data_view}
\begin{figure}[!ht]
    \includegraphics[width=.4\textwidth]{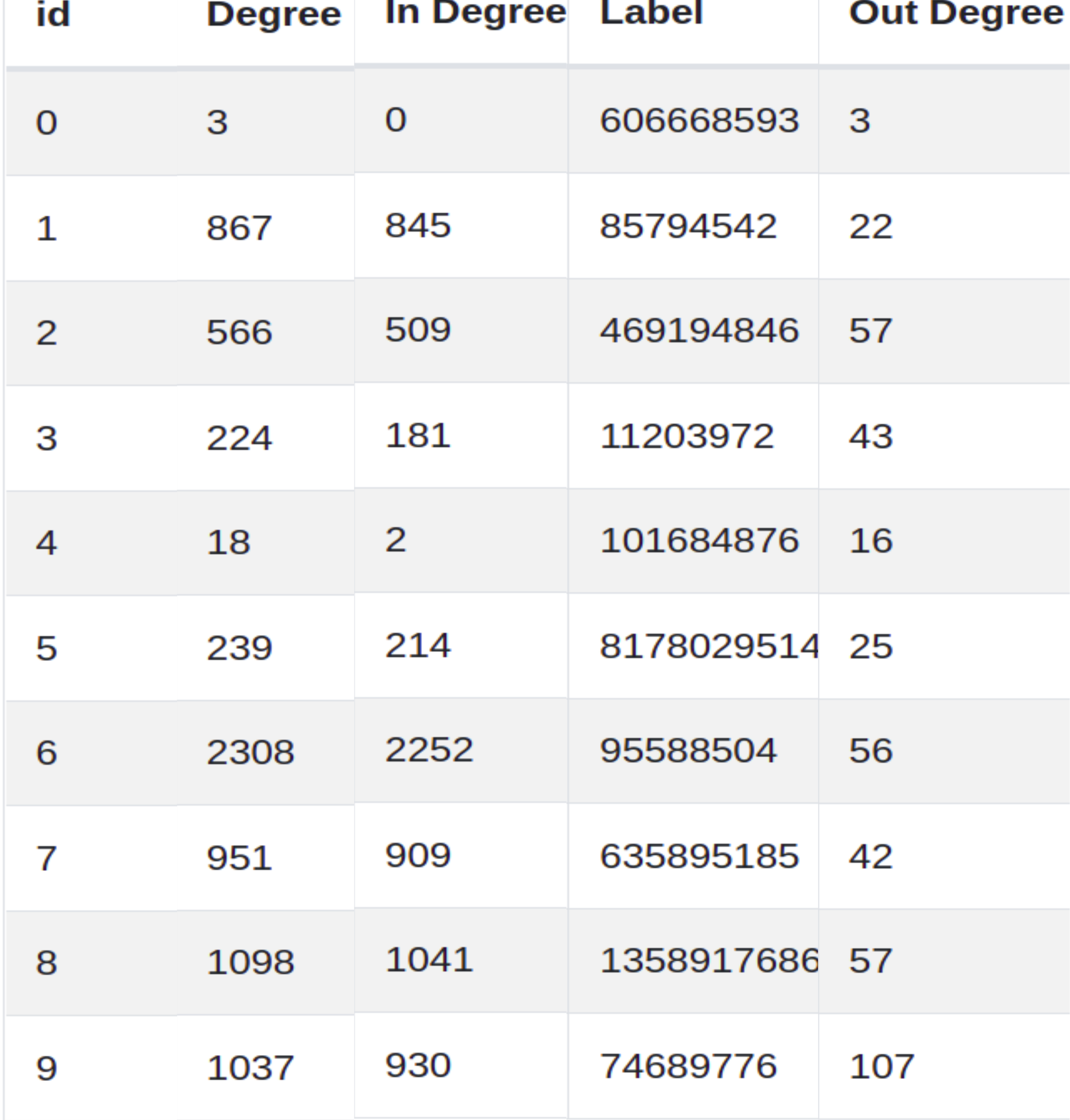}
    \caption{The \texttt{Data View} provides a columnar viewing of the current node attributes and statistics.}
    \label{fig:data_view}
\end{figure}

Once a network is loaded, users can click on the graph's nodes to view the details. However, to view label or attribute information about all the nodes, we have a different view format --- the \texttt{Data View}. It provides a columnar representation of all nodes at once. By default, it displays the following node attributes --- \textit{node-id, total degree, in/out-degree and label}. As and when the user runs more node level statistics, the respective columns appear in the \texttt{Data View}. Consequently, it provides the ability to search and sort by id and attributes. An example of this view is provided in Figure \ref{fig:data_view}.

\section{User Requirement Survey}
\label{app:req_survey}
Table \ref{tab:survey_req} enlists the questionnaire used for the user requirement survey. Note, for this survey users were not aware of the presence of our tool and they answered the questions based on their experience with other existing tools for network and diffusion analysis.
\begin{table}[]
\caption{Questionnaire for the user requirement survey.}
        \label{tab:survey_req}
\scalebox{0.75}{\begin{tabular}{p{0.28\textwidth}|p{0.38\textwidth}|p{0.3\textwidth}|p{0.2\textwidth}}
\hline
\textbf{Question Type} &
  \textbf{Question} &
  \textbf{Answer Format} &
  \textbf{Compulsory} \\
\hline
\multirow{4}{*}{Basic Information} &
  Please enter your name &
  Free text &
  \xmark \\ \cdashline{2-4}
 &
  Please enter your email address &
  Free text &
  \xmark \\ \cdashline{2-4}
 &
  How do you identify yourself? (Gender Orientation) &
  Free text &
  \xmark \\ \cdashline{2-4}
 &
  State your current occupation/role &
  Free text &
  \cmark \\ \hline
\multirow{8}{*}{Subject Knowledge} &
  What would you say is your level of expertise in the domain of network analysis? &
  Radio Button:1-5 (5 being the highest) &
  \cmark \\ \cdashline{2-4}
 &
  What do you understand by network diffusion? &
  Free text &
  \cmark \\ \cdashline{2-4}
 &
  Have you previously worked in the domain of network diffusion? &
  Radio Button: Yes/No &
  \cmark \\ \cdashline{2-4}
 &
  If "yes", which of the following have you used previously for visualization of network diffusion based analysis? &
  Check Box: Cytoscape, Gephi, Epinet Shiny App,  NDlib-Viz &
  \cmark \\ \cdashline{2-4}
 &
  If you use none of the above mentioned tools, how do you generally visualize diffusion dynamics on networks? &
  Free text &
  \xmark \\ \cdashline{2-4}
 &
  Can you list out some limitations that you may find while using some of the said tools? &
  Free text &
  \xmark \\ \cdashline{2-4}
 &
  According to your use cases, can you list 3 most important features that you feel must be present in any network diffusion visualisation tool that you may use? &
  Free text &
  \cmark \\ \cdashline{2-4}
 &
  On what platform would you prefer to use such an analytical tool? &
  Radio Button: Desktop application, Mobile/Tablet application, Web based interface, No preference &
  \cmark \\ \hline
\end{tabular}
}
\end{table}

\section{Complete Workflow}
\label{app:end2end_flow}
\subsection{Steps to Setup and Start \framework}
Before setting up \framework\ locally, the users are required to set up an app on the Google Developer Console\footnote{\url{https://console.cloud.google.com/}}. The system requires a Python-3 environment. The user can install the required python modules through the command \textpck{pip install -r requirements.txt}, and run the tool through \textpck{flask run}. Once the server starts, the user needs to open \textpck{http://localhost:5000} in a web browser. On the landing page, the user needs to authenticate via any google account and then click \textpck{Start Tool}. Note that in the case of using \framework\ as a hosted service, these steps can be forgone.  
    \subsection{Uploading Networks for Exploration}
        \begin{enumerate}
            \item When prompted with the \textpck{Setup Network} window, the user can generate a random graph using \textpck{Create Network} button after setting the necessary parameters, or upload an existing network via the \textpck{Upload Network}. Then, the {\em primary diffusion visualization} mode as described in Section \ref{sec:primary_viz_mode} will appear.
            \item To change the network's appearance, the user needs to go to the Left panel, \textpck{Appearance} $\hookrightarrow$ \textpck{Nodes} or \textpck{Edges} sub-tab.
            \item On the upper right panel, the network measures can be found under the \textpck{Context} $\hookrightarrow$ \textpck{Graph} sub-tab. 
            \item Under the \textpck{Statistics} section on the upper right panel, two sub-tabs are present --- \textpck{Network Overview}, and \textpck{Node Overview} to run network and node-level metrics, respectively. To view node level metrics, the user needs to select a node by clicking on it and switching over to the \textpck{Context} $\hookrightarrow$ \textpck{Selected Node} sub-tab in the upper Right panel.
            \item Alternatively, metadata for all nodes can be viewed under the \textpck{Data View} tab.
        \end{enumerate}
    \subsection{Running a Single Diffusion Model}
    \label{sec: single_model_steps_follow}
    Once the user is familiar with the network, they may want to run a few diffusion models to see how information spread occurs over this network. By default, the {\em primary diffusion visualization} mode is loaded.
        \begin{enumerate}
            \item The user can select the diffusion model, configure its parameters, set \textpck{Maximum Iteration} under the \textpck{Diffusion} tab in the upper left panel, and set the \textpck{Initial Infected Nodes} under the Initial Configuration in the bottom left panel. 
            \item After configuration, the user needs to click the \textpck{Submit} button under the \textpck{Diffusion} tab in the upper left panel. 
            \item At any point, the network's Appearance can be changed using the \textpck{Appearance} tab in the upper left panel.
            \item The user needs to use the \textpck{Play} button on the bottom panel to start visualizing the iterations. Alternatively, they can do it manually by pausing it (click on the $\rhd$ button again) and dragging the \textpck{Timeline slider}. \textpck{Next Frame} ($\gg$) button on the bottom panel moves the timeline forward by a one-time step.
            \item The user can view the diffusion statistics at each time step under the \textpck{Diffusion Statistics} tab in the upper right panel.
            \item The user needs to switch over to the \textpck{Report View} on the top panel to see a plot of diffusion trends. To download this report as a PDF, they can click on \textpck{Diffusion} $\hookrightarrow$ \textpck{Download Report}. 
            \item At any point, the user can seamlessly toggle between the \textpck{Report and Graph Views}.
        \end{enumerate}
    \subsection{Running Two Diffusion Models}
    In order to perform a comparative analysis between two different diffusion setups, the user needs to use the \textpck{Compare View} tab on the top panel, which will open a new window. The interface for the {\em dual diffusion visualization} mode is outlined in Section \ref{subsec:dual_diff}.
        \begin{enumerate}
            \item By default, the network will be loaded in the \textpck{Split View}.
            \item  Once the networks are loaded, two diffusion algorithms can be set in the \textpck{Diffusion} tab on the upper Left panel. Here, the user will be able to select a ground-truth result for comparative evaluation.
            \item Now the user can operate like Section \ref{sec: single_model_steps_follow}. The \textpck{Play} button, the \textpck{Timeline slider} and the \textpck{Report View} can be used in the same manner. Although, the enhanced \textpck{Diffusion Statistics} will now be available on the upper left panel.
            \item Here, the user has the option to switch to the \textpck{Single View} to visualize both algorithms working on a single network simultaneously.
            \item  The user needs to switch over to the \textpck{Report View} on the top panel to see a plot of diffusion trends, along with plots for commonly infected nodes and F1-score per iteration.
        \end{enumerate}
\end{document}